# The Impact of Different Haze Types on the Atmosphere and Observations of Hot Jupiters: 3D Simulations of HD 189733b, HD 209458b and WASP-39b

Mei Ting Mak,[1★] Denis Sergeev,[1,2] Nathan Mayne,[1] Maria Zamyatina,[1] Maria E. Steinrueck,[3†] James Manners,[4] Éric Hébrard,[1] David K. Sing,[5] Krisztian Kohary[1]

[1]*Department of Physics and Astronomy, University of Exeter, Exeter, UK*
[2]*School of Physics, University of Bristol, Bristol, UK*
[3]*Department of Astronomy and Astrophysics, University of Chicago, Chicago, IL, USA*
[4]*Met Office, Exeter, UK*
[5]*Department of Physics and Astronomy, Johns Hopkins University, Baltimore, MD, USA*

**ABSTRACT**

We present the results from the simulations of the atmospheres of hot-Jupiters HD 189733b, HD 209458b and WASP-39b, assuming the presence of three different types of haze. Using a 3D General Circulation Model, the Unified Model, we capture the advection, settling and radiative impact of Titan-like, water-world-like and soot-like haze, with a particle radius of 1.5 nm. We show that the radiative impact of haze leads to drastic changes in the thermal structure and circulation in the atmosphere. We then show that in all our simulations, 1) the superrotating jet largely determines the day-to-night haze distribution, 2) eddies drive the latitudinal haze distribution, and 3) the divergent and eddy component of the wind control the finer structure of the haze distribution. We further show that the stronger the absorption strength of the haze, the stronger the superrotating jet, lesser the difference of the day-to-night haze distribution, and larger the transit depth in the synthetic transmission spectrum. We also demonstrate that the presence of such small hazes could result in a stronger haze opacity over the morning terminator in all three planets. This could lead to an observable terminator asymmetry in WASP-39b, with the morning terminator presenting a larger transit depth than the evening terminator. This work suggests that, although it might not be a typical detection feature for hot-Jupiters, an observed increase in transit depth over the morning terminator across the UV and optical wavelength regime could serve as a strong indicator of the presence of haze.

**Key words:** planets and satellites: atmospheres – planets and satellites: gaseous planets – planets and satellites: composition – radiative transfer

## 1 INTRODUCTION

Photochemical hazes are small solid-state particles formed through photochemistry, suspended in an atmosphere (see e.g., Gao et al. 2021). They are found in the current atmosphere of Earth (Jimenez et al. 2009), and are expected to have been present on Earth 2.5 — 3.8 billion years ago during the Archean Eon (Zerkle et al. 2012). This is evidenced through the mass-independent fractionation of sulfur isotope in sedimentary minerals dating from the early Earth that hint at the presence of ultraviolet (UV) radiation absorbing species, potentially photochemical haze (Domagal-Goldman et al. 2008; Ueno et al. 2009; Zerkle et al. 2012; Liu et al. 2019). Beyond Earth, photochemical haze has been detected within our own Solar System. For example it was observed in the atmosphere of Titan by Pioneer 11, Voyager 1 and 2 and the Cassini/Huygens missions (Rages & Pollack 1980; Tomasko 1980; Rages et al. 1983; Rages & Pollack 1983; West et al. 1983; West 1991; West & Smith 1991; Tomasko et al. 2008, 2009). It was also detected in the atmosphere of Venus, for example by the Venus Express spacecraft and Akatsuki missions (Titov et al. 2018) and the Spectroscopy for Investigation of Characteristics of the Atmosphere of Venus/Solar Occultation at Infrared (SPICAV/SOIR) suite of instruments onboard the Venus Express (Wilquet et al. 2009, 2012; Luginin et al. 2016). Beyond the Solar System, observations such as from the Hubble Space Telescope (HST) and the James Webb Space Telescope (JWST) have shown a steep increase in the transit radius towards shorter wavelengths in the atmospheres of hot-Jupiters (Sing et al. 2016; Kempton et al. 2023), which are highly irradiated tidally-synchronised gas giants (Dole 1964; Kasting et al. 1993; Guillot et al. 1996; Barnes 2017). This spectral feature indicates scattering from small particles, potentially photochemical haze (Sing et al. 2009; Nikolov et al. 2015; Sing et al. 2016; Wong et al. 2020; Ohno & Kawashima 2020; Spake et al. 2021; He et al. 2024).

When haze is present in the atmosphere of a planet, it can alter the radiative forcing, influencing the atmospheric structure and circulation. However, only limited studies have been performed which explore the impact of this radiative feedback of haze on the climate. For terrestrial exoplanets, Arney et al. (2016) coupled the 1D pho-tochemical Atmos model, based on the network from Pavlov et al. (2001), with a 1D climate model to simulate the impact of Titan-like haze on the Archean Earth. Arney et al. (2016) found that an optically thin haze layer almost has no radiative impact on the climate. Whereas an optically thick haze layer results in strong cooling in the lower atmosphere due to the strong shortwave heating in the upper atmosphere, also referred to as the anti-greenhouse effect. The results from Arney et al. (2016) are in agreement with other 1D work (Pavlov et al. 2001; Haqq-Misra et al. 2008; Zerkle et al. 2012). However, performing climate simulations using 3D General Circulation Models (GCMs) is important to accurately capture the radiative impact of haze on the thermal structure and atmospheric transport, resolving complex features like jet streams and overturning circulation which are inherently 3D. Considering this, Mak et al. (2023) and Mak et al. (2024) prescribed a Titan-like haze distribution from the Atmos model for the Archean Earth and TRAPPIST-1e, respectively, within a 3D GCM to simulate their radiative impact. At odds with previous 1D work, Mak et al. (2023, 2024) showed that when prescribing a thin haze layer in the atmosphere of terrestrial planets, it led to surface warming due to the radiative forcing from haze changing the water vapour distribution and cloud area fraction.

★ E-mail: mtm206@exeter.ac.uk
† 51 Pegasi b fellow

For gas giant exoplanets, current full photochemical models include over 4000 chemical reactions (Venot et al. 2015), but are limited to 1D configurations due to the high computational cost in calculating their time evolution. They do not include the formation of haze from photochemistry, but only the formation of haze precursors. Morley et al. (2015) performed calculations using a photochemical network applied to the atmosphere of the sub-Neptune, GJ 1214b, to obtain the total mass of various soot-like haze precursors. Soot-like haze refers to the carbonaceous particles formed in combustion experiments commonly used when simulating haze in hot-Jupiters (Morley et al. 2015; Lavvas & Koskinen 2017; Lavvas & Arfaux 2021; Steinrueck et al. 2021, 2023; Arfaux & Lavvas 2024). Morley et al. (2015) then used a 1D radiative-convective model to simulate the impact of soot-like haze, showing a strong shortwave heating and significant increase of temperature in the upper atmosphere, similar to the results seen with thick haze layer in the atmosphere of terrestrial planets discussed above.

In order to make study of haze production due to photochemistry more computationally feasible, a fixed particle mass flux has been applied to simulations previously (Lavvas & Koskinen 2017; Ohno & Kawashima 2020; Lavvas & Arfaux 2021; Arfaux & Lavvas 2022; Steinrueck et al. 2021, 2023; Gao et al. 2023; Arfaux & Lavvas 2024). Lavvas & Arfaux (2021) estimated the particle mass flux based on the photolysis of haze precursors. They then used a 1D radiative-convective model to simulate the impact of Titan-like and soot-like haze in the atmosphere of HD 189733b. Lavvas & Arfaux (2021)'s work also show the anti-greenhouse effect of haze, and highlighted the importance of using 3D GCMs to study the radiative impact of haze on the entire atmospheric circulations in gas giants. Steinrueck et al. (2021, 2023) developed a parameterisation within the 3D MITgcm, assuming a fixed haze production profile following a log-normal distribution in pressure in the atmosphere of HD 189733b. Their study also involved Titan-like and soot-like haze with a mass flux ranging between $1\times10^{-10}$–$2.5\times10^{-12}$ kg m$^{-2}$ s$^{-1}$. Steinrueck et al. (2023) show that in addition to strong heating in the upper atmosphere, the type of haze also influences the formation of different jet structures and strengths. Steinrueck et al. (2023) and He et al. (2024) also suggested that varying haze types and optical properties can impact the transmission spectra of planets differently. Given that equilibrium temperatures of hot-Jupiters ranges between 1000–2000 K, the haze formed may well be very different across this population (Khare et al. 1984; Trainer et al. 2004, 2006; He et al. 2018; Hörst et al. 2018; Fleury et al. 2019, 2020; He et al. 2020, 2022, 2024). Additionally, different hot-Jupiters exhibit varying atmospheric circulations (Showman et al. 2009), potentially bringing haze particles to different locations which will change the shortwave absorption of these haze particles. Therefore, further study is required on how various haze properties and distributions impact the resulting planetary atmospheric dynamics and observations of different hot-Jupiters. This work is particularly important given the quality and volume of hot-Jupiter observations available, or planned from JWST.

In this work we couple the Unified Model (UM), a 3D GCM developed by the UK Met Office, to a haze production and removal parameterisation and study the impact of varying haze types on three of the best-studied hot-Jupiters, namely, HD 189733b, HD 209458b and WASP-39b. We apply Titan-like, water-world-like and soot-like haze as indicative of haze produced on rocky planets, sub-Neptunes and hot-Jupiters respectively, and explore their impact on the atmospheric dynamics and the synthetic spectra of the target hot-Jupiters. We find that the radiative impact of haze leads to significant changes in the thermal structure, which subsequently change the atmospheric circulation. While the precise changes in circulation and haze distribution vary from planet to planet due to their unique atmospheric dynamics, all simulations show that 1) the superrotating jet largely determines the day-to-night haze distribution, 2) eddies drive the latitudinal haze distribution, and 3) the divergent and eddy component of the wind control the finer structure of the haze distribution. In particular, our results show that the stronger the absorption strength of the haze, the stronger the superrotating jet, lesser the difference of the day-to-night haze distribution, and larger the transit depth in the synthetic transmission spectrum. The rest of this paper is laid out as follows: Sec. 2 describes our haze and climate model. Sec. 3 presents the results from the simulations. Sec. 4 discusses the implications of our results and the comparison with previous work. Finally, Sec. 5 summarizes the key findings of this paper. The ultimate goal of this paper is to provide a broader picture detailing the effect of haze on the atmospheric circulation and observations of hot-Jupiters.

## 2 MODEL

In this work we combine a photochemical haze production treatment, with the adoption of three types of well characterised haze optical properties within a 3D GCM, the Unified Model (UM). We then perform simulations of the planets HD 189733b, HD 209458b and WASP-39b for all three types of haze, and for cases where the haze is radiatively passive (without the radiative feedback from haze on the thermal structure) and active (with the radiative feedback from haze on the thermal structure). In this section we first describe the haze production (Sec. 2.1) and sources of optical data (Sec. 2.2), before detailing the 3D simulations we have performed (Sec. 2.3).

### 2.1 Haze Model

We adapt the haze model of Steinrueck et al. (2021, 2023) where the haze mass mixing ratio (MMR; defined as the mass of haze per mass of air) $\chi$ is calculated using,

$$\frac{D\chi}{Dt} = -g\frac{\partial(\rho_a \chi V_s)}{\partial p} + P + L \quad , \tag{1}$$

where $D/Dt$ is the material derivative, $g$ is the gravitational acceleration, $t$ is the time, $p$ is the pressure, $\rho_a$ is the mass density of air and $V_s$ is the Stokes velocity. The first term on the right describes the settling of haze particles. $P$ and $L$ are the haze production and loss term, respectively. $P$ is assumed to be a log-normal distribution in pressure given by,

$$P = F_0 g \cos\theta \frac{1}{\sqrt{2\pi} p\sigma} \exp\left(-\frac{(\ln(p/m))^2}{2\sigma^2}\right) \quad (2)$$

where $F_0$ is the column-integrated haze mass production rate with the value of $1\times10^{-12}$ kg m$^{-2}$ s$^{-1}$, $\theta$ is the zenith angle of the incident radiation, and $m$ and $\sigma$ are the median pressure and geometric standard deviation of the distribution in pressure. Steinrueck et al. (2021) and Steinrueck et al. (2023) adopted $m$ and $\sigma$ values of 0.002 mbar and 0.576, respectively. They have chosen these distribution parameters such that the width of their haze production profile becomes negligible for the top two layers in their 3D GCM model setup, MITgcm. In our simulations we do not extend to pressures as low as those captured in Steinrueck et al. (2023). Instead we adopt an adjusted value for $m$ of 0.005 mbar to aid model stability (discussed in Sec. 2.3), but retain the $\sigma$ of 0.576. We note that this downward shift does not lead to significant differences compared to the results from Steinrueck et al. (2023) after a pseudo-steady state has been reached (see further discussions in Secs. 2.3 and 4.3). Additionally, since the primary focus of this work is to investigate the radiative impact of haze on the atmospheric dynamics of different hot-Jupiters, and how that would subsequently impact observable features, we have opted to configure the haze model in a way that prioritises numerical stability, rather than replicating the exact same haze production profile as used in previous work (see further discussions in Sec. 4.1).

The loss term $L$ takes into account the removal of haze particles by thermal destruction and condensation of cloud species onto haze particles. $L$ is given by,

$$L = \begin{cases} -\chi/\tau_{\rm loss} & p > p_{\rm deep}, \\ 0 & \text{elsewhere}, \end{cases} \quad (3)$$

where $\tau_{\rm loss}$ is the loss timescale, and $p_{\rm deep}$ sets the boundary for haze removal. Both of which are set to the values of Steinrueck et al. (2021, 2023) for our simulations, namely, $10^3$ s and 100 mbar, respectively. Here we note that the results presented in Sec. 3 are sensitive to the choice of model parameters. For instance, adopting a lower value of $F_0$ would reduce the haze concentration, therefore weakening the radiative forcing within the atmosphere and altering the outcomes of the simulations. We also note that this haze model is a simplified framework that assumes a fixed haze production profile, omitting potential variation in the haze formation environment (see further discussions in Sec. 4.1). Yet, again the primary goal of this work is to explore how haze influences atmospheric circulation and observable spectral features. As a result, a simplified approach is opted for this study to isolate the radiative effects of haze within a controlled setting. The implementation of a fully self-consistent chemical network will be needed for future work.

### 2.2 Haze Types

Three types of haze are used in this work: Titan-like, water-world-like and soot-like. In all cases, haze particles are assumed to be spherical with a log-normal radius distribution, centred at 1.5 nm and a geometric standard deviation of 1.5. The assumption of spherical haze particles allows the adoption of Mie scattering (Bohren & Huffman 2008) to calculate their optical properties of haze particles (see Sec. 2.3 for more details). We note that the adoption of haze particles with such small sizes implies that they are more susceptible to advection than to gravitational settling. In other words, they are more prone to staying aloft than settling to the deeper atmosphere due to gravity. Our findings therefore may not be representative of scenarios involving larger haze particles in the atmosphere (see further discussions in Sec. 4.2).

For Titan-like haze, we adopt the same optical properties used in Mak et al. (2023) and Mak et al. (2024), which studied the impact of haze on the Archean Earth and TRAPPIST-1e, respectively. For this case, haze particles are assumed to have a mass density of 640 kg m$^{-3}$ (Arney et al. 2016; Mak et al. 2023, 2024). The optical properties are taken from Khare et al. (1984), which covers the wavelength range of 0.027–920 $\mu$m, but replaced with the updated values of He et al. (2022), in the wavelength range of 0.4–3.5 $\mu$m. The haze samples from both Khare et al. (1984) and He et al. (2022) were generated by subjecting a mixture of $N_2$ and $CH_4$ to a cold plasma discharge. For water-world-like haze, we adopt the optical properties of the haze samples generated by He et al. (2024). To study the potential haze formation in a water-rich exoplanet atmosphere, He et al. (2024) generated thin films of haze by subjecting a mixture of $H_2O$ (the dominating species), $CH_4$, $CO_2$, $N_2$, $H_2$ and He to a cold plasma discharge at 400 K. These gas mixtures were calculated under chemical equilibrium, representing an atmosphere with 1000× solar metallicity (see He et al. (2024) for further details). These water-world-like haze particles have a mass density of 1262 kg m$^{-3}$, and optical properties covering 0.4–28.6 $\mu$m, and Khare et al. (1984) for the wavelength that lies outside of that wavelength range but are needed as input for the radiative transfer code in the GCM (details discussed in Sec. 2.3). Finally, for soot-like haze, we follow Steinrueck et al. (2021), Steinrueck et al. (2023), Lavvas & Arfaux (2021), Arfaux & Lavvas (2024), Kempton et al. (2023) and Gao et al. (2023) and adopt the optical properties of Lavvas & Koskinen (2017). Lavvas & Koskinen (2017) combines results from Lee & Tien (1981), Chang & Charalampopoulos (1990) and Gavilan et al. (2016) who generated their haze samples from flame combustion experiments of hydrocarbons. Their experimental results cover the wavelength range of 0.268–228 $\mu$m. Following Steinrueck et al. (2021) and Steinrueck et al. (2023), the soot's particle mass density of 1000 kg m$^{-3}$ is adopted. We note here that the haze type does not alter the production and loss term described in Sec. 2.1 (see Eqns. 2 and 3).

The real ($n$) and imaginary ($k$) components of the refractive indices, along with the extinction efficiency and single scattering albedo calculated through Mie theory, of all three types of haze simulated in this work are compared in Fig. 1. Fig. 1(i) shows that all three hazes have a similar refractivity at short wavelengths. However, soot-like haze has a greater refractivity at longer wavelengths. Figs. 1(ii–iv) shows that the absorption, extinction efficiency and single scattering albedo of soot-like haze are less dependent on wavelength (i.e., a grey optical profile) but its absorption is $10^1$–$10^2$ times higher than the other two types. In general, water-world-like haze is slightly less absorbing than Titan-like haze and has a weaker extinction efficiency, except in the wavelength range between 0.5–1.1 $\mu$m. We note that the Titan-like haze from Khare et al. (1984) and He et al. (2022) is produced at room temperature and ~100 K, respectively. Whereas, the water-world-like haze from He et al. (2024) is produced at ~400 K. Although these laboratory environments are likely different to those present in the high temperature atmospheres of hot-Jupiters, the data required to fully constrain the type and properties of hazes present in hot-Jupiters are not currently available. Therefore, in this work, to aid in our understanding of the haze impacts and properties we perform simulations adopting the three types of haze for which we can obtain data for at the time of writing.

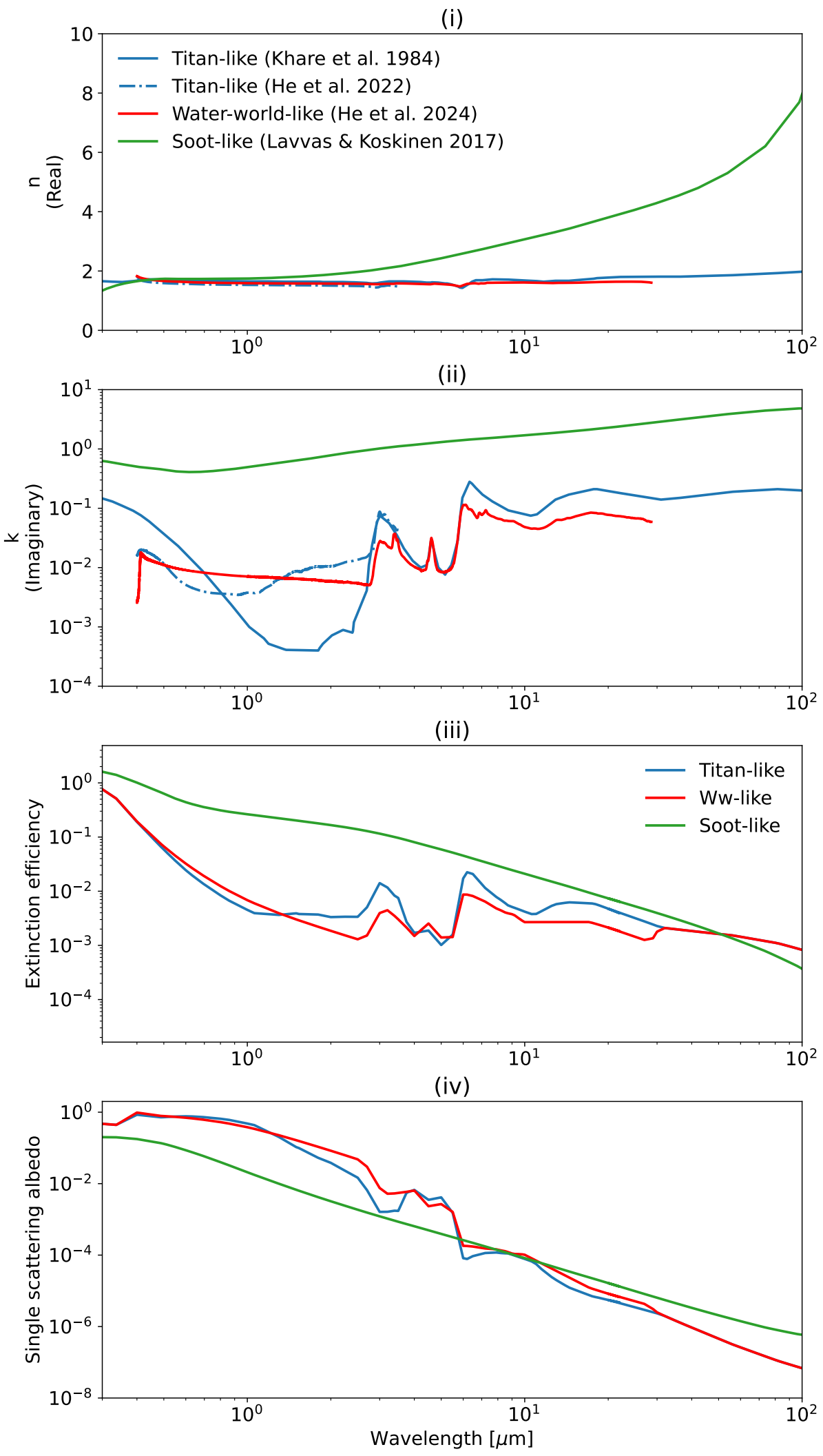

**Figure 1.** Comparison of complex refractive indices (i–ii), extinction efficiency (iii) and single scattering albedo (iv) from Titan-like (Khare et al. 1984; He et al. 2022), water-world-like (He et al. 2024) and soot-like haze (Lavvas & Koskinen 2017). The extinction efficiency (iii) and single scattering albedo (iv) are calculated from the Mie theory using SOCRATES (see details in Sec. 2.3).

## 2.3 3D Simulations

We use the Unified Model (UM), the 3D GCM developed by the UK Met Office, to simulate the impact of haze on our target planets. The dynamical core of the UM, ENDGame (Even Newer Dynamics for General atmospheric modelling of the environment), uses the semi-implicit semi-Lagrangian scheme to solve the non-hydrostatic, full deep-atmosphere equations of motion in the atmosphere with varying gravity (see Wood et al. 2014; Mayne et al. 2014a,b, 2017, 2019, for discussion). The UM has been adapted to study modern Earth (Walters et al. 2019; Andrews et al. 2020), the Archean Earth (Eager-Nash et al. 2023; Mak et al. 2023), Mars (McCulloch et al. 2023), terrestrial exoplanets (Boutle et al. 2020; Sergeev et al. 2022; Ridgway et al. 2023; Mak et al. 2024) and gas giant exoplanets (Mayne et al. 2019; Christie et al. 2021, 2022; Zamyatina et al. 2023, 2024).

The stellar and planetary parameters adopted for each our of targets, HD 189733b, HD 209458b and WASP-39b are listed in Tab. 1 and Tab. 2, respectively. The stellar and planetary parameters for HD 189733b and HD 209458b are taken from the TEPCat data base [1] and Southworth (2010). The stellar and planetary parameters used for WASP-39b are taken from GAIA Data Release 3 (DR3) [2] and Mancini et al. (2018) respectively, unless stated otherwise. The solar abundances are taken from Asplund et al. (2009). The alkali metal abundances included in our work use the parameterisation from Amundsen et al. (2016), in which they apply additional smoothing to the analytical fit of the alkali metals' monatomic/polyatomic transformation boundaries taken from Burrows & Sharp (1999). The specific gas constant in each planetary atmosphere is calculated from the initial temperature and chemical equilibrium profile based on the metallicity in Tab. 2. Such value has also been adopted in previous work, with HD 189733b in this paper sharing the same value with Mayne et al. (2014b) and Heng et al. (2011), and HD 209458b with Drummond et al. (2018b).

A horizontal grid spacing of 2.5° in longitude and 2° in latitude is used in all simulations within this study. The substellar point is set at 180° longitude and 0° latitude. The UM uses a height-based vertical coordinate and we use 80 uniformly spaced vertical levels for HD 209458b and WASP-39b, and 70 for HD 189733b. The choices of model domain height and number of vertical levels in all cases are a balance between numerical stability and attempting to obtain approximately similar vertical resolutions in pressure. For all three planets, the pressure of the bottom boundary is $2\times10^5$ mbar. The dayside atmosphere heats from the initial condition, leading to an increase in pressure at the upper model boundary. To aid numerical stability, we iterate all simulations to achieve the lowest pressure at the upper boundary on the dayside. At pseudo-steady state, the dayside pressure at the top-of-atmosphere reaches ~0.15 mbar for HD 189733b ($4.5\times10^6$ m, see Tab. 2), ~0.001 mbar for HD 209458b ($11\times10^6$ m, see Tab. 2), and ~0.05 mbar for WASP-39b ($4.5\times10^6$ m, see Tab. 2). As discussed in Sec. 2.1, we set the median pressure of the haze production distribution to 0.005 mbar, meaning that the peak of haze production (Eq. 2) lies at lower pressures than our dayside upper boundary for HD 189733b and WASP-39b. As a result, haze production is capped at this altitude and any production occurring at lower pressures is not captured. However, this capping has little effect on the final haze concentration once the simulation has reached a pseudo-steady state. This is supported by comparison with Steinrueck et al. (2023), whose model domain extends below the peak haze production region (see Section 4.3 for further discussions).

The radiative transfer calculation within the UM is performed by the "Suite Of Community RAdiative Transfer codes based on Edwards & Slingo (1996) (Socrates). Socrates is a two-stream radiative transfer code that uses the correlated-$k$ method to solve the gaseous absorption from $H_2O$, $CH_4$, CO, Cs, K, Li, Na, $NH_3$, Rb and collision-induced absorption from $H_2$–$H_2$ and $H_2$–He from the ExoMol line lists (Tennyson et al. 2016). The stellar spectrum used for HD 189733 and HD 209468 are generated from the Kurucz model [3], whereas the stellar spectrum used for WASP-39 is generated from the PHOENIX BT-settl model (Rajpurohit et al. 2013). Socrates is used to construct an input file for the UM (termed a 'spectral file') containing the normalised stellar flux, gaseous absorption and the extinction, scattering and asymmetry coefficients of the haze particles across 32 spectral bands covering a wavelength range of 0.2–323$\mu$m. Socrates is also capable of calculating synthetic transmission spectra

[1] https://www.astro.keele.ac.uk/jkt/tepcat/
[2] https://www.cosmos.esa.int/web/gaia/dr3
[3] http://kurucz.harvard.edu/stars.html

**Table 1.** Stellar parameters used in the three model configurations.

| Stellar parameter | HD 189733 | HD 209458 | WASP-39 |
|---|---|---|---|
| Type | K1-K2 | G0 | G8 (Faedi et al. 2011) |
| Radius [m] | $5.2\times10^8$ | $7.0\times10^8$ | $6.5\times10^8$ |
| Effective temperature [K] | 5050 | 6100 | 5512.0 |
| Stellar constant at 1 AU [W m$^{-2}$] | 447.0 | 2054.7 | 905.3 |
| $\log_{10}$(surface gravity) [cgs] | 4.53 | 4.38 | 4.47 |

**Table 2.** Planetary parameters used in the three model configurations.

| Planetary parameter | HD 189733b | HD 209458b | WASP-39b |
|---|---|---|---|
| Inner radius [m] | $8.1\times10^7$ | $9.0\times10^7$ | $8.9\times10^7$ |
| Domain height [m] | $0.45\times10^7$ | $1.1\times10^7$ | $2.5\times10^7$ |
| Semi-major axis [AU] | 0.03142 | 0.04747 | 0.04828 |
| Orbital period [Earth day] | 2.2 | 3.5 | 4.1 |
| Rotation rate [rad s$^{-1}$] | $3.3\times10^{-5}$ | $2.1\times10^{-5}$ | $1.8\times10^{-5}$ |
| Surface gravity [m s$^{-2}$] | 21.5 | 9.3 | 4.3 |
| Metallicity | 1×solar | 1×solar | 1×solar |
| Specific gas constant [J K$^{-1}$ kg$^{-1}$] | 4593 | 3557 | 3516 |
| Specific heat capacity [J K$^{-1}$ kg$^{-1}$] | $1.3\times10^4$ | $1.3\times10^4$ | $1.3\times10^4$ |

for the UM simulations (for details see Sec. 4.8 in Lines et al. 2018; Villanueva et al. 2024).

For simplicity, we have not included UV photolysis (Baeyens et al. 2022), thermal chemistry (Cooper & Showman 2006; Venot et al. 2012; Tsai et al. 2017; Tsai et al. 2022; Drummond et al. 2018a,b; Venot et al. 2019, 2020; Drummond et al. 2020; Zamyatina et al. 2023, 2024) or clouds (Lee et al. 2016; Lines et al. 2019; Christie et al. 2021, 2022) both to increase the feasibility of the study and also isolate the effect of haze. We have run the simulations for at least 1200 Earth days so that a pseudo-steady state is reached in the upper atmosphere. This is determined by the change of the haze distribution, upward wind velocity and balance of the global top-of-atmosphere net radiative flux being $\leq$1%. The last 50 Earth days are temporally averaged for all data analysis, except for the calculation of synthetic transmission spectra in which we have used the data from the last time step of the simulation. We note that the deeper atmosphere is still evolving (see Sainsbury-Martinez et al. 2019, for discussion). The bottom few layers of our simulations for WASP-39b exhibit small signs of numerical instability. To aid interpretation, smoothing has been applied in the affected region. However, we stress that these instabilities do not affect the overall conclusions of this study. Furthermore, our analysis focuses on the upper atmosphere at $p\leq10$ mbar (all simulations have reached a pseudo-steady state beyond this pressure level), which corresponds to the pressure levels where the relevant observations of interest to this paper are made. Additionally, regions at higher pressures show smaller differences compared to the pressure region of 10 mbar and are therefore not the focus of this study.

## 3 RESULTS

We begin by detailing the haze distribution of the passive haze cases, demonstrating how the circulation determines the spatial distribution of the haze. Afterwards, we introduce radiative feedback from the haze, and present the resulting differences in the simulations from the passive cases. Finally, we present synthetic transmission spectra for all of our targets and qualitatively compare them to observational data. We label simulations with radiatively passive haze with “[p]” and the active haze with “[a]”.

### 3.1 Radiatively Passive Haze

In this subsection, we begin by detailing the small-particle haze distributions for the cases of radiatively passive hazes ([p]) where there is no radiative feedback from haze on the thermal structure or dynamics. We then show that the superrotating jet determines the (day-to-night) side haze distribution, whereas the eddies determine the latitudinal variation of haze distribution as demonstrated by the eddy mass flux. Afterwards, we introduce a subsection for each planet where we show that the decomposed element of the atmospheric circulation determines the finer details and specific haze distribution pattern.

We start by examining the global haze distribution across pressure levels. The haze production term stated in Eqn. 2 describes a production rate increasing towards lower pressures, peaking at 0.005 mbar. Additionally, this assumed production term depends on the cosine of the incident angle, $\cos\theta$. This is because the haze model assumes maximum haze production at the substellar point, where the stellar flux is normal to the surface and the irradiation is strongest. As a result, without the consideration of atmospheric circulation, one would expect the haze MMR, on the day side, to increase as 1) the pressure decreases, and/or 2) the latitude decreases, as well as expecting 3) depletion of haze on the night side.

Fig. 2 shows the zonal mean haze distributions as a function of pressure for all cases of HD 189733b, HD 209458b and WASP-39b. For the passive haze cases, Figs. 2(i–iii) show that, in general, all three planets exhibit a higher haze MMR towards lower pressures. However, Figs. 2(i–iii) also show that the haze MMR peaks at mid-latitudes as opposed to at the equator where the $\cos\theta$ term is largest. Furthermore, Fig. 3, which shows the dayside- and nightside-mean haze profiles, reveals that not only is haze present on the night side in passive haze cases ([p]), but is also showing higher haze MMR at all pressure levels over the night side than on the day side for all three planets. Fig. 3 also shows that the peak haze MMR occurs at ap-proximately the same pressures on both day and night side, although for HD 189733b and WASP-39b this is at the top boundary (see discussion in Sec. 2.3), whereas for HD 209458b the haze MMR peaks before remaining approximately constant with decreasing pressure. The presence of haze on the night side where there is no incident radiation, and the fact that the haze distributions do not peak at the equator indicate that advection in addition to production and loss (see Eq. 1) shape the haze distribution.

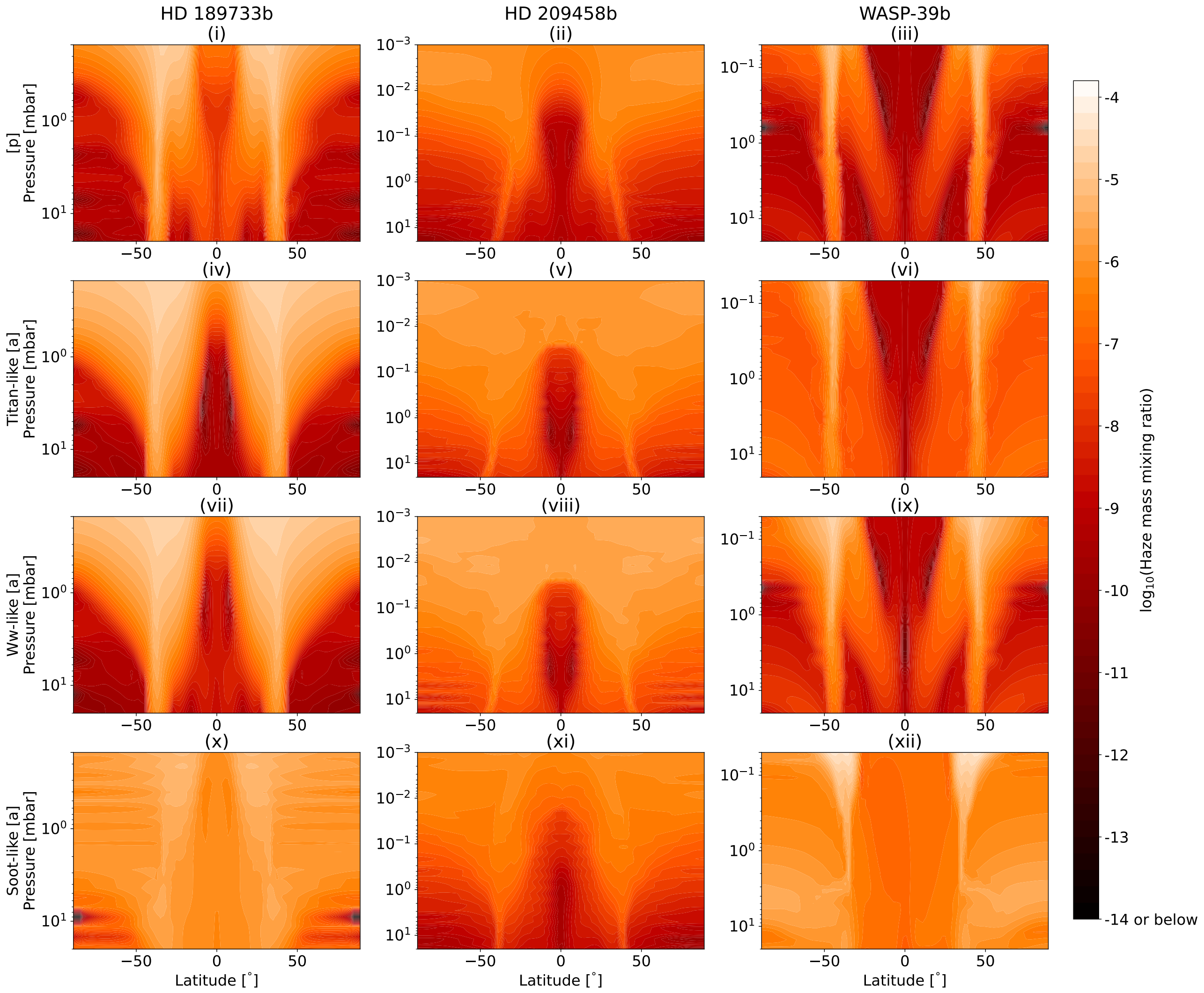

**Figure 2.** Zonal mean of haze mass mixing ratio as a function of latitude and pressure for all cases of HD 189733b, HD 209458b and WASP-39b. The haze mass mixing ratio is expressed in $\log_{10}$ scale. "[p]" and "[a]" represent the passive and active haze case, respectively. "Ww" stands for water-world.

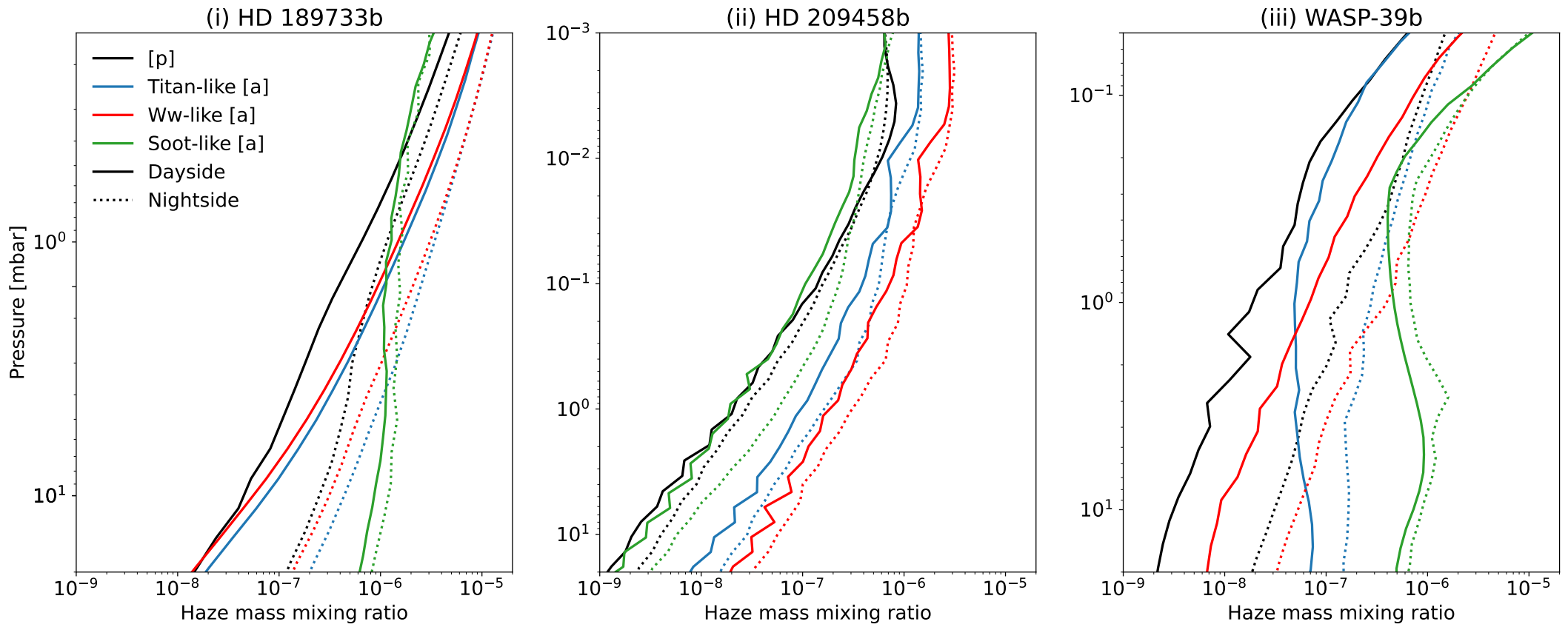

**Figure 3.** Dayside- (solid) and nightside-mean (dotted) haze mass mixing ratio profiles for all cases of HD 189733b, HD 209458b and WASP-39b. "[p]" and "[a]" represent the passive and active haze case, respectively. "Ww" stands for water-world.

To understand the large-scale day-to-night pattern of haze distribution, Fig. 4 shows the zonal mean zonal wind velocity as a function of latitude and pressure for all cases, with contours showing the haze MMR on a logarithmic scale. Only $log_{10}$(haze MMR)>-10 is plotted here to ease interpretation. Focusing on the passive haze cases (i.e., [p], Figs. 4(i-iii)), the zonal flows of all three planets are dominated by a prograde superrotating jet at the equator. The strong jet in all three planets allows the transport of haze from the day side (where haze production is taking place) towards the night side, permitting the build-up of haze particles in both hemispheres as shown in Fig. 3.

Figs. 4(i-iii) also show that in the upper atmosphere, the peak haze MMR generally aligns with the regions where the prograde flow transitions to retrograde, or simply the "edges" of the jet. This correlation can be understood through the eddy mass flux

$$\frac{\partial v'\chi'}{\partial \phi}, \quad (4)$$

where $v$ is the meridional velocity, $\phi$ is the latitude and the prime denotes a deviation from the zonal mean: $A' = A - \bar{A}$. A positive eddy mass flux corresponds to regions where haze is being transported to by the eddies, whereas regions with a negative eddy mass flux shows where the eddies are removing haze. Fig. 5 shows the zonal mean eddy mass flux, as a function of latitude and pressure, with contours of haze MMR also shown following the same format as Fig. 4. Focusing on the passive cases at lower pressure levels, Figs. 5(i-iii) show that the zonal mean eddy mass flux is strongly positive in regions correlating, approximately, of haze accumulation and the edge of the superrotating jet. As pressure increases, the eddy mass flux becomes weaker around the edge of the jet, reducing the concentration of haze in such regions.

To summarise, our results from the small-particle passive haze cases show that 1) the day-to-night haze distribution is driven by the superrotating jet, while 2) the latitudinal distribution is driven by the eddies as shown by the eddy mass flux. The investigation of the specific haze distribution over these planets requires separate analysis of the planet's own circulation pattern, which will be discussed in the following subsections.

#### *3.1.1 HD 189733b [p]*

Fig. 6 shows the latitudinal and longitudinal variation of haze MMR for all cases of HD 189733b at three isobaric surfaces, namely pressures of 0.15, 1 and 10 mbar. Focusing on the passive haze cases (i.e., [p]; Figs. 6(i–iii)), Fig. 6(i) shows that at 0.15 mbar, haze concentrates at ±20° around the edge of the jet (see Fig. 4(i)), peaking over the night side at ±40° latitude. As pressure increases, haze remains concentrated around the night side at the same latitude, but its distribution at the edge of the superrotating jet becomes relatively lower.

Fig. 7 shows the wind field of HD 189733b for all haze types, for the same isobaric surfaces as Fig. 6. Focusing solely on the passive haze cases (i.e., [p], Figs. 7(i–iii)) for 0.15 mbar, the peak vertical velocity is present in the upwelling region on the day side, acting to keep the haze particles aloft. The strong eddies, as demonstrated by the positive eddy mass flux in Fig. 5(i), further allow the build-up of haze at the edge of the jet. To explore further the impact of the wind structure, we can study each of the wind components and disentangle the effect from large-scale circulation and eddies. Fig. 8 shows the Helmholtz decomposition of the wind at 0.15 mbar for all cases of HD 189733b, following Hammond & Lewis (2021). Focusing on the passive cases (i.e. [p], Figs. 8(i–iii)), Fig. 8(iii) shows strong cyclonic eddies on the night side close to the morning terminator, also known as the nightside vortices, which act to trap haze particles (Steinrueck et al. 2021, 2023; Zamyatina et al. 2024). These nightside vortices also locate very close to the edge of the jet (see Figs. 8(ii–iii)). This wind structure results in the haze MMR shown in Fig. 6(i), peaking in these regions.

At higher pressures, the structure of the decomposed wind closely resembles the pattern shown in Figs. 8(i–iii) and is therefore not plotted here. It again exhibits cyclonic eddies around the night side, which facilitate the build up of the haze particles around the nightside vortices. However, with a smaller eddy mass flux (see Fig. 5(i)), less haze is trapped around the edge of the jet, leaving a majority of haze concentrate around the nightside vortices only.

#### *3.1.2 HD 209458b [p]*

Fig. 9 shows the haze MMR for all cases of HD 209458b, following the format of Fig. 6 but at isobaric surfaces of pressure 0.001, 0.01, 0.1 and 1 mbar. For the passive haze cases at a pressure of 0.001 mbar, Fig. 9(i) shows that haze is concentrated on the day side from the mid-latitude to polar regions, peaking over the evening terminator at ±50° latitude. On the night side, the haze also concentrates around ±40° latitude at the edge of the jet. The region where the morning terminator intersects with the equator is relatively devoid of haze, compared to the rest of the atmosphere. As the pressure increases to 0.01 mbar, the haze is mostly concentrated only at ±30° latitude around the edge of the jet peaking over the anti-stellar longitude. As the pressure increases further, the haze concentrates mostly over the nightside vortices at mid-latitudes, similar to HD 189733b.

Fig. 10 shows the wind field for all cases of HD 209458b for pressures matching those of Fig. 9. Fig. 11 then shows the wind decomposition for the 0.001 mbar isobaric surface for all cases of HD 209458b. Focusing only on the passive cases, Figs. 11(i and iii) show that a strong divergent and eddy component help distributing the haze efficiently towards the mid-latitude to polar regions. In particular, the anticyclonic eddies eastward of the day side bring more haze towards the evening terminator.

As the pressure increases, the structure of the decomposed winds are again similar to those shown, and are not plotted here. However, as discussed above from Figs. 9(ii-iv), haze is mostly concentrated around the edge of the jet at ~30° peaking over the anti-stellar longitude, with a comparatively lower haze concentration in the mid- to high-latitude area, as opposed to what is seen at pressure = 0.001 mbar. This can be explained by the positive eddy mass flux locating at lower latitudes, allowing the build-up of haze there. In the meantime, the negative eddy mass flux at higher latitudes indicates that the eddies are transporting haze away (see Fig. 5(ii)). Therefore less haze is found from mid-latitude to polar regions on the day side. As the pressure increases even further, the eddy mass flux decreases. As a result the hazes particles are trapped over the nightside vortices only (see Fig. 5(ii)), very similar to what is seen in HD 189733b.

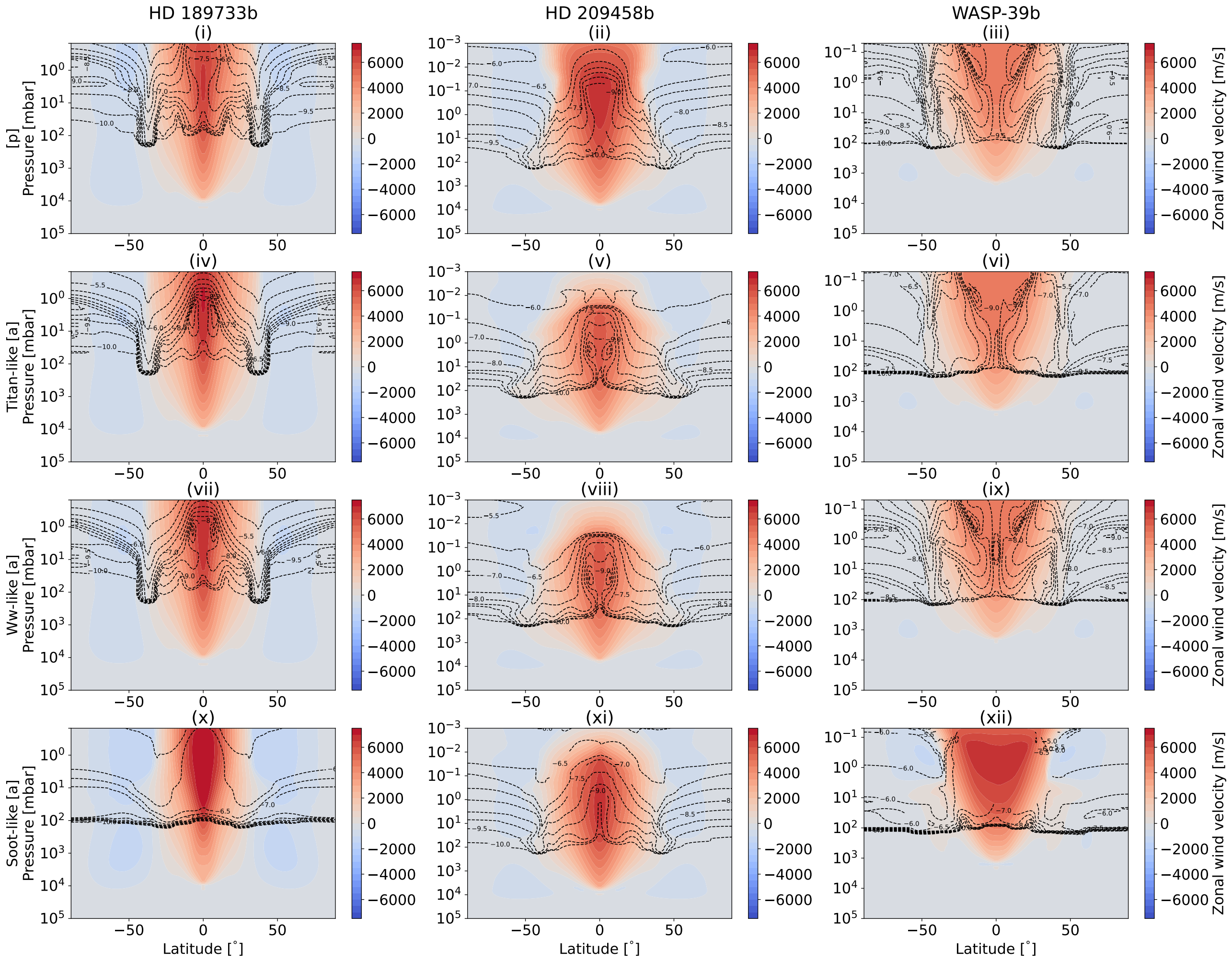

**Figure 4.** Zonal mean zonal wind velocity, as a function of latitude and pressure, with contours showing the haze MMR in logarithmic scale for all cases of HD 189733b, HD 209458b and WASP-39b. Only $log_{10}$(haze MMR)>-10 is shown here. “[p]” and “[a]” represent the passive and active haze case, respectively. “Ww” stands for water-world.

### 3.1.3 *WASP-39b [p]*

Fig. 12 shows the distribution of haze for all cases of WASP-39b. For the passive haze case at a pressure of 0.05 mbar, Fig. 12(i) shows that haze again concentrates around the edge of the jet and mid- to high-latitudes, peaking over the nightside vortices at ±50° latitude. The haze is relatively depleted from the equator to mid-latitude compared to rest of the atmosphere. As pressure increases, haze distribution remains similar, resembling to what is observed in HD 189733b.

Figs. 13 and 14 show the wind field and the decomposition of the wind, respectively, for WASP-39b. For the passive haze case at a pressure of 0.05 mbar, Fig. 13(i) shows that the equatorial upwelling region over the dayside is slightly weaker than those in higher latitudes. The haze particles therefore sink to deeper layers, resulting in a lower haze concentration between the equator to mid-latitude. The strong eddies, as shown by the positive eddy mass flux in Fig. 5(iii), again results in the concentration of haze around the edge of the jet. Fig. 14(iii) shows that the strong cyclonic eddies help trap haze over the nightside vortices. As pressure increases, the structure of the decomposed winds at these pressure levels is again similar, resulting in a very similar distribution across pressure levels.

To summarise, for all three simulated planets, even though radiatively passive small-particle haze is produced on the day side where the atmosphere is receiving stellar radiation, the equatorial superrotating jet transports the haze particles from the day to night side. The eddies transport the haze to the edges of the jet, bringing the haze away from the equator to mid-latitude, as demonstrated by the eddy mass flux distribution.

## 3.2 Radiatively Active Haze

In this subsection, we present the simulation results where we include the radiative impact of Titan-like, water-world-like and soot-like haze (labelled with [a] in all figures). We first discuss the heating rates due to the presence of different hazes and their effect on the subsequent thermal structure. We show that soot-like haze heats up the atmosphere the most, changing thermal structures significantly compared to the other haze cases. We then present changes in the global distribution of haze, superrotating jet and eddy mass flux due to these changes in the thermal structure, showing that the day-to-night and latitudinal haze distribution are still determined by the superrotating jet and eddies (as shown through the eddy mass flux), respectively. Afterwards, we introduce a subsection for each planet where we show the changes to the decomposed element of the atmospheric circula-tion due to the radiative impact of haze, explaining the specific haze distribution observed in each case.

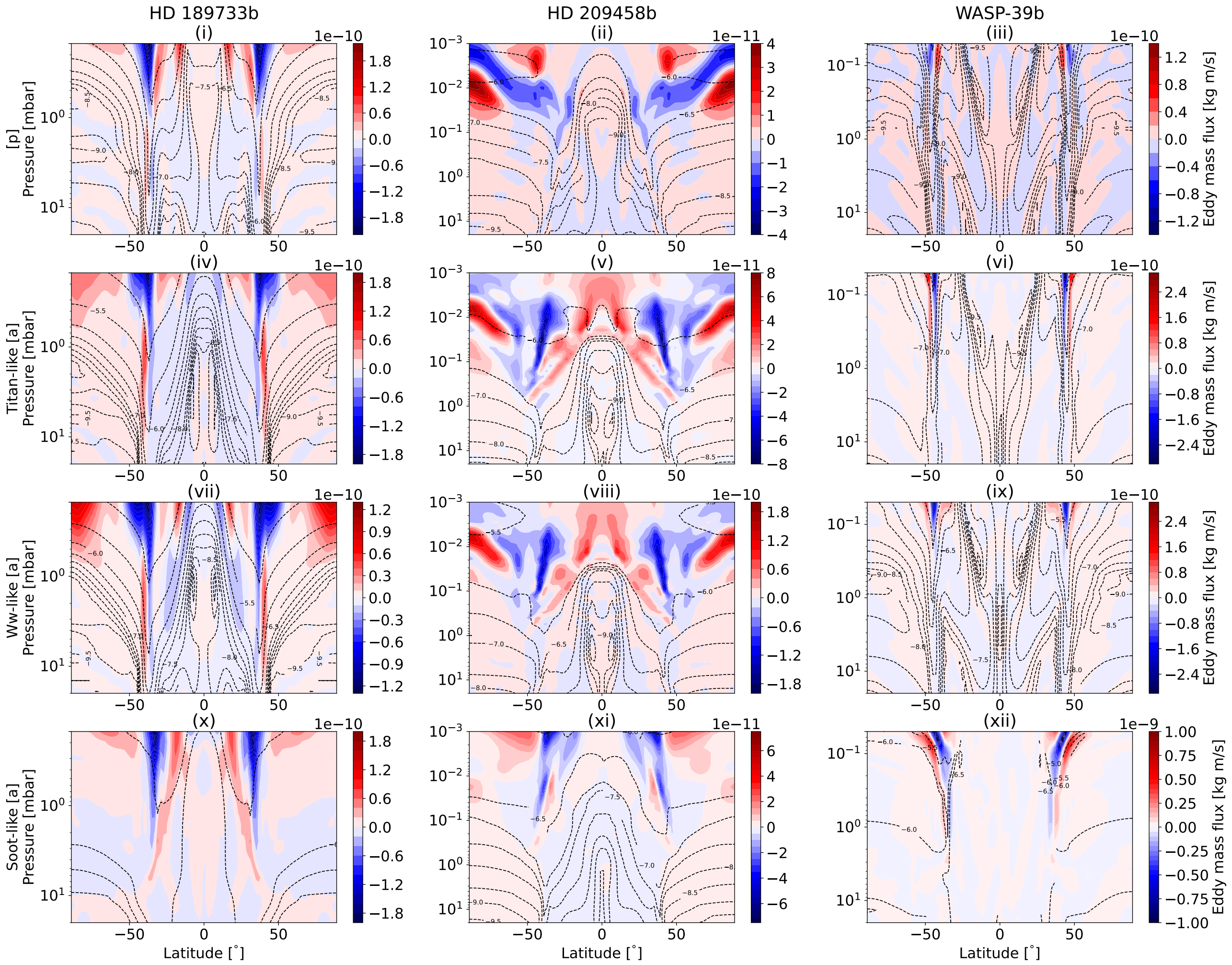

**Figure 5.** Zonal mean eddy mass flux, as a function of latitude and pressure, with contours showing the haze MMR in logarithmic scale for HD 189733b, HD 209458b and WASP-39b, following the format of Fig. 4. Definition of eddy mass flux is given in Eqn. 4.

As the UM is a height based model, at the top of the atmosphere the pressure decreases sharply moving from the day to night side. This gradient, alongside the strong shortwave heating from haze can lead to spatial fluctuations in the temperature, wind field and haze distribution close to the upper boundary. This effect is particularly pronounced when a planet experiences a strong shortwave heating, notably for the active soot-like haze cases in all three planets. For this work we make note of this fluctuation and focus on general trends. Additionally, in the upper atmosphere of the soot-like haze case from HD 189733b and HD 209458b, we have taken the mean value between the neighbouring model levels when presenting the vertical profiles. This modification is applied to Figs. 3, 17, 19, and 20.

To understand the differences in the results between different types of hazes, it is important to understand the radiative effect of haze. Fig. 15 shows the dayside-mean heating rate as a function of pressure. Figs. 15(i–iii) show that for all cases and for all planets, the soot-like haze causes the strongest shortwave heating rate. This is due to the absorption strength in the soot-like haze being the strongest (see Fig. 1). The heating rate from Titan-like and water-world-like haze share a similar trend, with the water-world-like haze showing a slightly weaker heating rate. From Fig. 1 and as discussed in Sec. 2.2, water-world-like haze shows a similar optical profile to Titan-like haze and therefore the two would also share a similar heating profile in all three planets.

With soot-like haze as the strongest absorber, simulations of all planets including this type of haze present the strongest jet, with the core of the jet pushed deeper into the atmosphere, compared to other active haze cases (see Fig. 4(xii) and Tab. 3). The cause for the deeper and stronger jet formation can be understood from the location of the net radiative heating. Figs. 16(i–iii) show the dayside-mean net heating rate of all planets. For HD 189733b and WASP-39b (panel (i) and (iii) respectively), soot-like haze exhibits the strongest net heating rate across all pressure levels considered here (i.e., ≤10 mbar). For HD 209458b (panel ii), soot-like haze exhibits the strongest net heating rate for pressures of 0.002–0.03 mbar and ≥2 mbar. As heating in the atmosphere is one of the key drivers of global circulation, increased heating would result in a stronger jet (see Figure 4 from Koll & Komacek (2018)). The stronger net heating for the soot-like case in the deeper atmosphere therefore results in the formation of the stronger jet at higher pressures than for the other active haze cases (for further discussion see Sec. 4.3).

Fig. 17 shows the dayside- and nightside-mean thermal structure as a function of pressure. Fig. 17 shows that for all planets, the presence of haze leads to an increase in temperature for both the day and night side. The thermal structure under the presence of Titan-like and water-world-like haze are very similar, and the simulations with soot-like haze show the highest increase of temperature, as expected (see Figs. 1, 15 and 16). Due to the haze-induced shortwave heating, an isothermal region and thermal inversion in the thermal structures are observed for all our simulations (except for the Titan-like and water-world-like haze cases in HD 209458b where only isothermal structure is observed). These changes to the thermal structure drive a different circulation pattern, resulting in a different haze distribution.

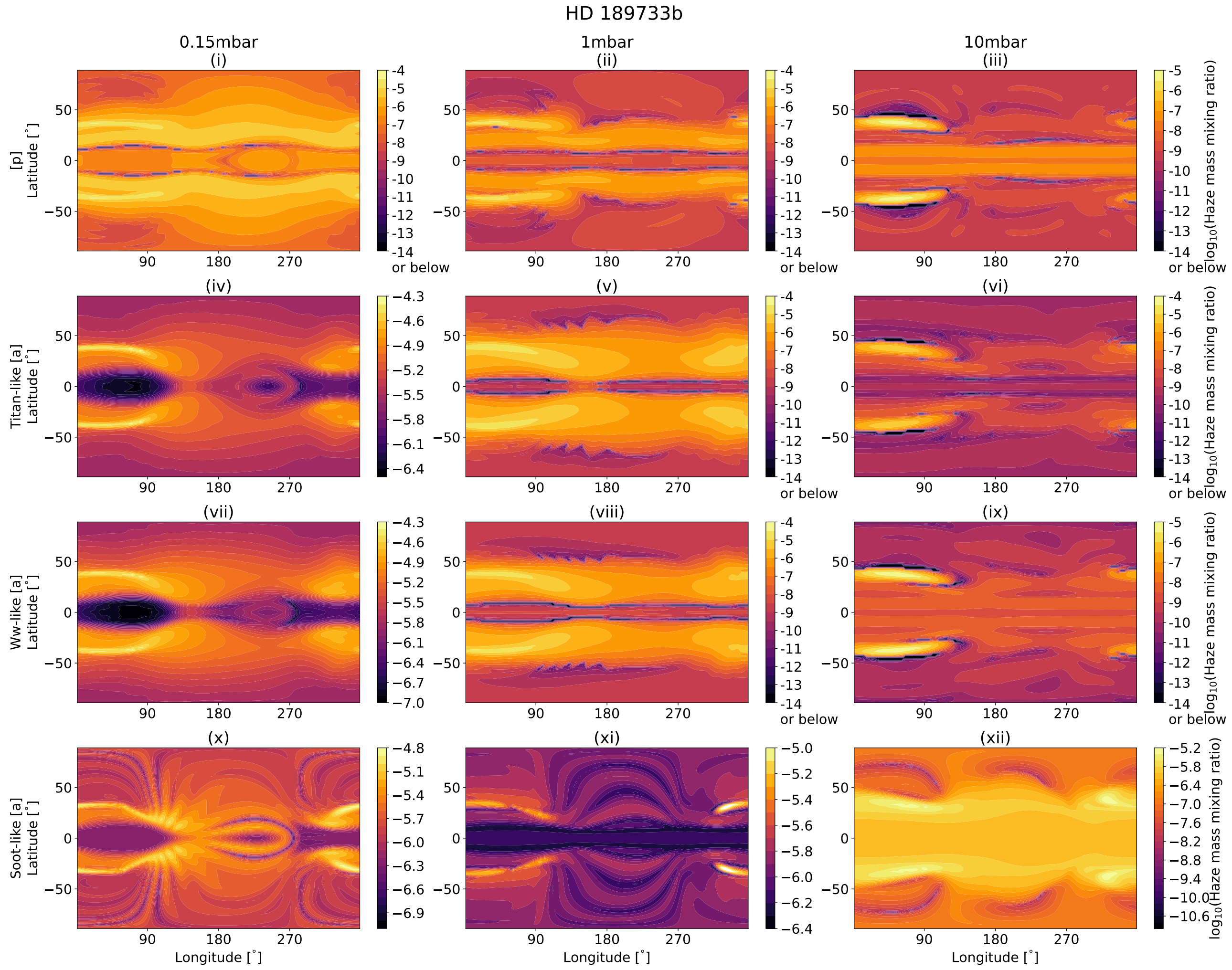

**Figure 6.** Mass mixing ratio of haze of HD 189733b for all cases at three isobaric surfaces, namely pressures of 0.15, 1 and 10 mbar. The haze mass mixing ratio is expressed in $\log_{10}$ scale. Note that the colour scales are different between different panels. "[p]" and "[a]" represent the passive and active haze case, respectively. "Ww" stands for water-world.

Figs. 3(i–iii) show that in all active haze cases there is haze on the night side. This demonstrates that the equatorial superrotating jet continues to efficiently carry haze particles from the production region on the day to night side regardless of the radiative forcing (see Figs. 4(iv–xii)). This matches the conclusions drawn in Sec. 3.1. Figs. 3(i–iii) also show that the stronger the absorption strength of the haze, the smaller the difference in the day-to-night haze distribution. This is due to the formation of a stronger jet, facilitating day-to-night mixing of haze. The details of the atmospheric circulation will be discussed specifically for each planet in Secs.3.2.1, 3.2.2 and 3.2.3.

Figs. 2(iv–xii) show that the zonal mean haze MMR in all active haze cases peaks in the mid-latitude regions at lower pressure, similar to that in the passive haze cases. This latitudinal distribution can again be explained by the positive zonal mean eddy mass flux at mid-latitudes as shown in Figs. 5(iv–xii). However, Figs. 5(iv–xii) also show that the eddy mass flux structure changes due to the radiative impact of haze, and for the case of HD 209458b, the positive eddy mass flux mid-latitude regions no longer correlate with the edge of the jet. This means that the positive eddy mass flux is driving the accumulation of haze towards higher latitudes but is no longer around the edge of the jet (details will be discussed in Sec. 3.2.2).

To summarise, including the radiative impact of haze significantly changes the thermal structure and alters the atmospheric circulation. Our results show that the stronger the absorption strength of the haze, the stronger the superrotating jet and lesser the difference in the day-to-night haze distribution. However, our simulations with radiatively active hazes generally agree with the passive haze cases, demonstrating that the same mechanisms drive the small-particle haze distribution, i.e., the superrotating jet determines the day-to-night haze distribution while the eddies determine its latitudinal haze distribution. To investigate how the radiative impact of each haze type influences the details of circulation pattern, and how that would in turn determine the exact haze distribution, we analyse the atmospheric dynamics for each planet under the presence of different hazes in the following subsections.

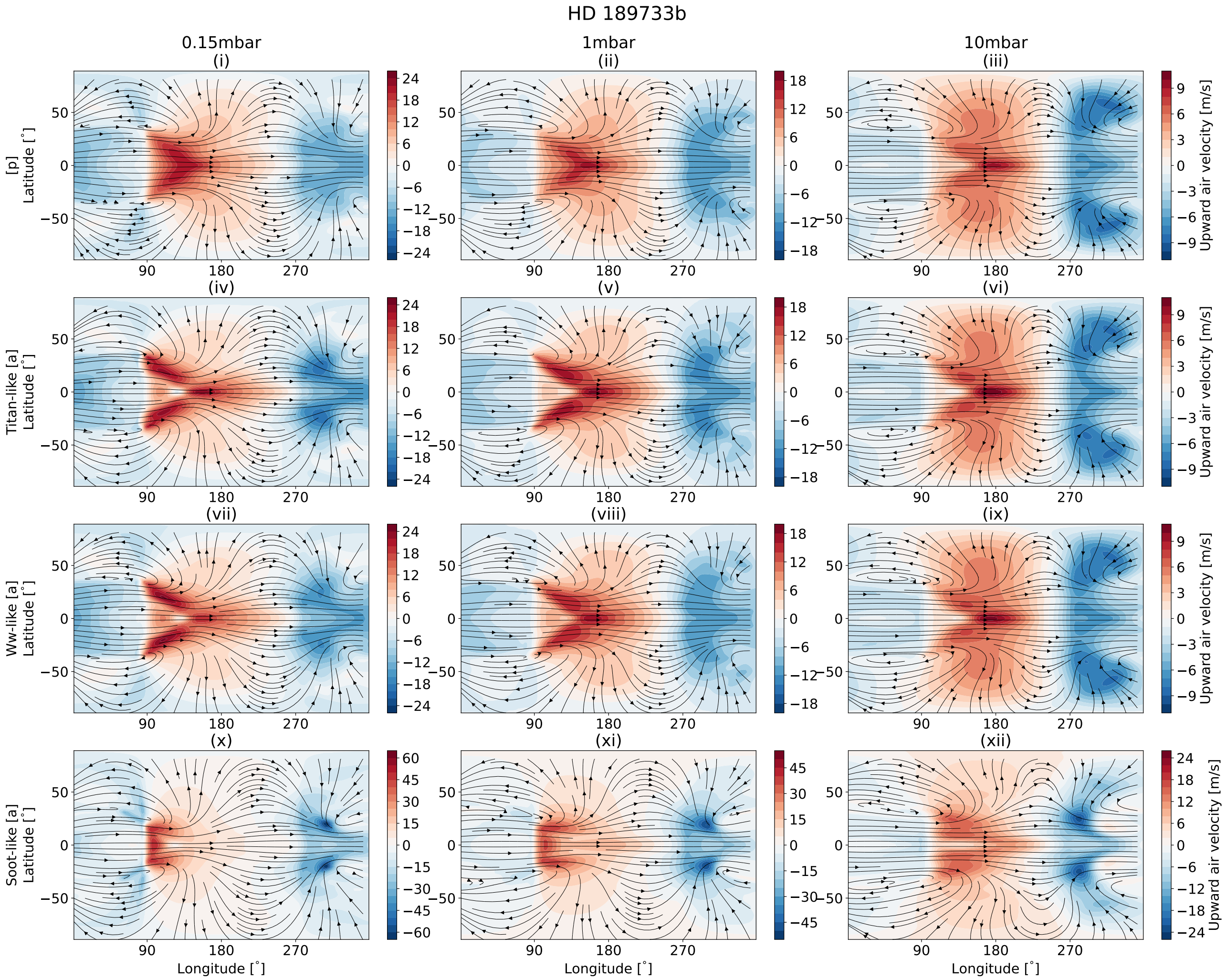

**Figure 7.** Vertical air velocity (contours) and horizontal wind (arrows) from simulations of HD 189733b, following the format of Fig. 6.

### 3.2.1 *HD 189733b [a]*

For the upper atmosphere of HD 189733b, Figs. 6(iv–xii) show that all active haze cases in general share a similar distribution of haze, which concentrates around the edge of the superrotating jet at ±40° latitude (see Figs. 4(iv, vii and x)), peaking in the nightside vortices and overlapping with the antistellar longitude. Note that the soot-like haze case exhibits spatial fluctuations in the haze distribution near the top of the atmosphere, as explained at the beginning of Sec. 3.2 (see Figs. 6(x–xi)). The Titan-like and water-world-like haze cases reach a maximum MMR over the nightside vortices. Whereas the soot-like haze case reaches a maximum MMR over the antistellar longitude. The MMR distributions from all of the active haze cases resemble the passive haze case. However, in these active haze cases, the haze distribution at ∼135° longitude extends closer to the equator, which is not observed in the passive haze case. As pressure increases, the haze distribution in the Titan-like and water-world-like haze cases remains concentrated in the nightside vortices with a smaller latitudinal variation, similar to the passive haze case. However for the soot-like haze case, not only does the haze deposit around the night side vortices, but remain high in concentration around the jet (see Fig. 6(xii)).

Figs. 7(iv, vii and x) show that at a pressure of 0.15 mbar, the vertical wind structure of the Titan-like and water-world-like haze cases are similar to the passive haze case. They have a slightly stronger downwelling velocity over the night side near the evening terminator compared to the passive haze case. They also present a more prominent chevron-shape upwelling region. However the centre of the chevron-shape upwelling region (near the equator at ∼135° longitude) exhibits almost zero upward velocity, which is not seen in the passive haze case. The soot-like haze case exhibits a different vertical wind structure to the other cases. The strong shortwave heating from the soot-like haze results in the chevron-shaped upwelling region being no longer visible for the day side, unlike the other two active haze cases. Yet, the zero vertical velocity region near the equator at ∼135° longitude is still present.

Even though the radiative effect of the haze is changing the vertical structure differently in these active haze cases, their horizontal wind structures are similar, resulting in a similar haze distribution. In particular, at pressure of 0.15 mbar, Figs. 5(iv, vii and x) show that the strong positive eddy mass flux is again correlated with the edge of the jet, allowing the build-up of haze particles there, similar to the passive haze case. Figs. 8(iv, vii and x) also show that a stronger convergence is seen at the equator between the morning terminator and ∼135° longitude compared to the passive haze case, bringing the haze particles from the edge of the jet to the converging point.

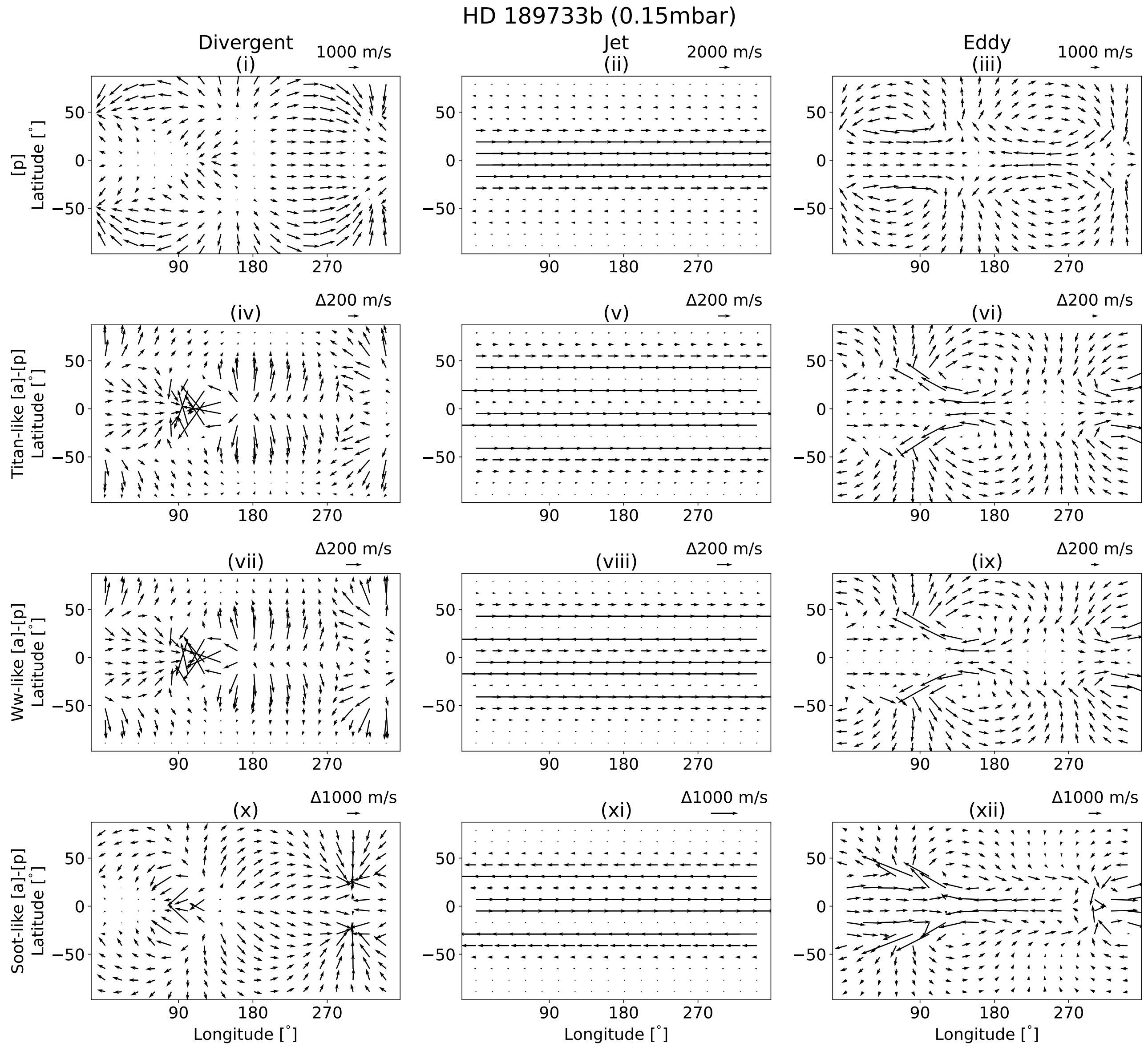

**Figure 8.** Divergent (left column), jet (middle column) and eddy (right column) component of the horizontal wind at a pressure of 0.15 mbar for all cases of HD 189733b. The active haze cases are presented as the difference (indicated by Δ) of the wind velocity of the active haze case minus that of the passive haze case. "[p]" and "[a]" represents the passive and active haze case, respectively. "Ww" stands for water-world.

This change in horizontal wind pattern and haze distribution is not seen in the passive haze case. The minor difference between the three active haze cases is that the eddy strength for the soot-like haze case is much stronger than the other two active haze cases (see Figs. 8(vi, ix and xii)). The soot-like haze case shows stronger cyclonic eddies across the night side over the antistellar longitude. As a result, this soot-like haze case also shows the concentration of haze around the antistellar longitude at pressure of 0.15 mbar (see Fig. 6(xi)).

At higher pressures, the wind structure for the Titan-like and water-world-like haze cases is similar to the passive haze case which results in a similar haze distribution pattern (see discussion in Sec. 3.1.1). For the soot-like haze case, the positive eddy mass flux remains strong in the deeper atmosphere (see Fig. 5(x)), allowing the haze particles to continue to be trapped near the edge of the jet at a higher pressure level, unlike the other cases. Additionally, Figs. 7(xi–xii) show that the night side of the planet is also exhibiting an upwelling region, which is not seen in other cases. This keeps the haze particles aloft over the night side at these pressure levels, appearing over the nightside vortices overlapping with the antistellar longitude.

### 3.2.2 *HD 209458b [a]*

Figs. 9(v–xvi) show that all three active haze cases in general share a similar haze distribution, similar to what is seen in HD 189733b. Note that the soot-like haze case also exhibits spatial fluctuations in the haze distribution as shown in Figs. 9(xiii). At pressure of 0.001 mbar, both Titan-like and water-world-like haze cases exhibit regions of haze at mid-latitude over the night side, transitioning to the day side through the equator. On the day side, they show regions of haze distribution near the pole. In particular, the Titan-like and water-world-like haze cases show a maximum haze MMR located over the morning terminator around ±10° latitude. For the soot-like haze

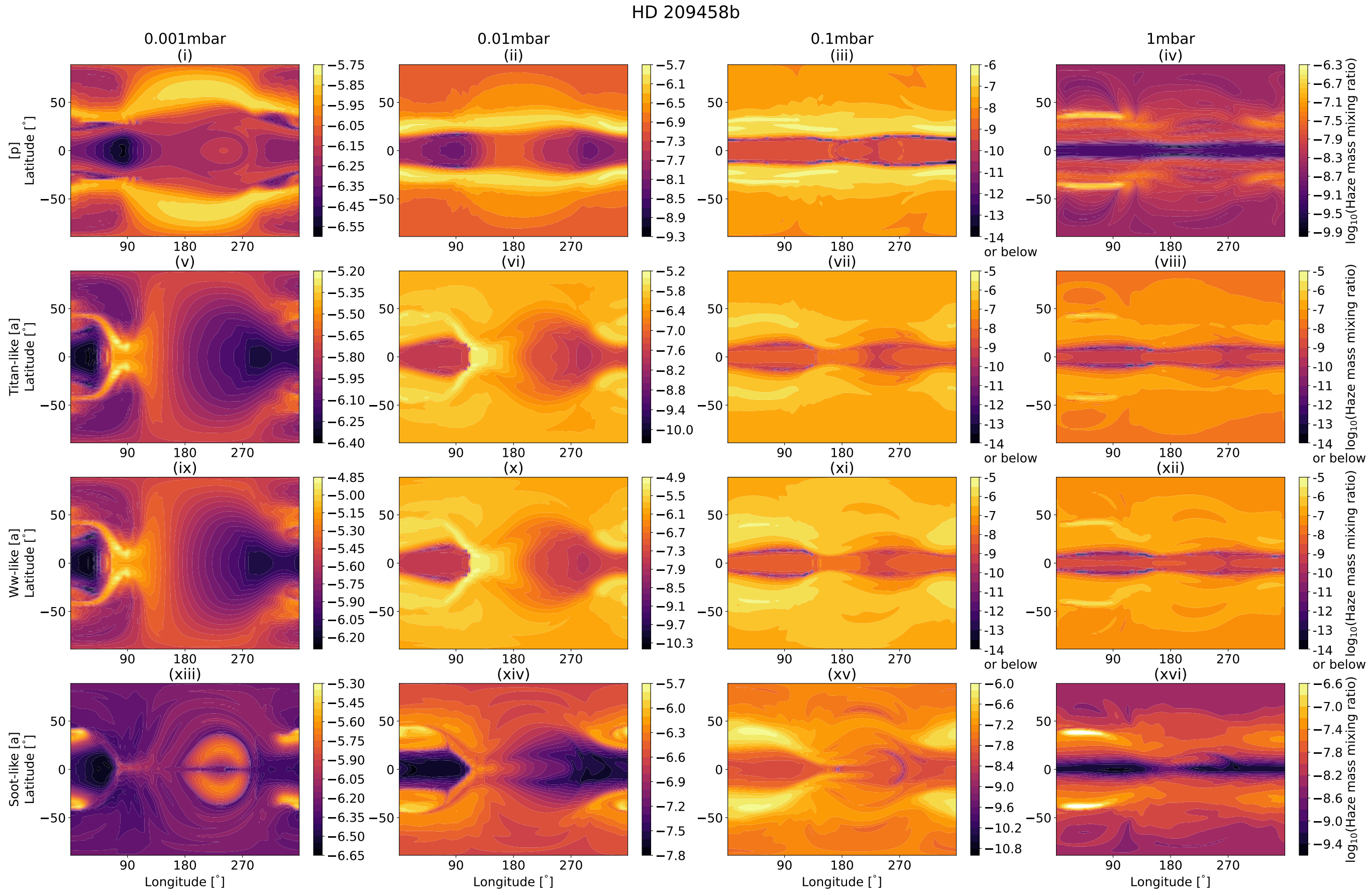

**Figure 9.** Mass mixing ratio of haze of HD 209458b for all cases at four isobaric surfaces, namely pressures of 0.001, 0.01, 0.1 and 1 mbar. The haze mass mixing ratio is expressed in $\log_{10}$ scale. Note that the colour scales are different between different panels. "[p]" and "[a]" represents the passive and active haze case, respectively. "Ww" stands for water-world.

case, haze instead concentrate over the nightside vortices overlapping with the antistellar longitude. There is also relatively more haze in a circular region eastward of the dayside near the equator compared to other regions, which is not seen in the other two active haze cases. All of these distribution patterns do not resemble the passive haze case (see Fig. 9(i)). At higher pressures, the haze partly covers the nightside vortices and the edge of the jet, resembling the passive haze case and the active haze case for HD 189733b.

Figs. 10(v, ix and xiii) show that at a pressure of 0.001 mbar, the vertical wind structure of all active haze cases are similar. All of them show a very weak upward velocity on the day side, in comparison with strong downwelling region at mid-latitude over the morning terminator. Breaking down the wind field into its components, Figs. 11(iv, vii and x) show that the flow is highly divergent at the substellar point, creating westward flow on the day side towards the morning terminator, compared with the passive haze case.

Furthermore, Figs. 11(vi, ix and xii) show that comparing to the passive haze case, stronger eddies converge at the equator at ∼50° longitude, then move towards the mid-latitude for both hemispheres. These changes to the wind component overall introduce a strong converging flow (see Figs. 10(v, ix and xiii)), bringing the haze particles to the morning terminator across the equator around ±10° latitude for the Titan-like and water-world-like haze case. The converging eddies also explain the strong positive eddy mass flux at ±10° latitude from Figs. 5(v and viii). For the soot-like haze, Fig. 10(xiii) shows that there is weaker vertical velocity on the day side, compared to the other two cases. Therefore, despite the wind diverging under the substellar point and converging towards ∼50° longitude and the equator like the other two active haze cases, a lesser amount of haze is observed at the converging point. Instead, comparatively more haze is concentrated over the antistellar longitude around ±40° latitude, matching the strong positive eddy mass flux at the same location from Fig. 5(xi).

For all three active haze cases at higher pressure (0.1–1 mbar) towards the centre of the jet, the diverging and converging flow are diminishing (see Figs. 10(vii-viii, xi–xii and xv–xvi)). The upwelling region begins to show the chevron-shaped pattern, with more haze trapped over the nightside vortices. Positive eddy mass flux is also formed (see Figs. 5(v, viii and xi)), bringing a small amount of haze to the edge of the jet again. The simulations at higher pressure level are similar to the passive haze case, as well as the results from HD 189733b.

### 3.2.3 *WASP-39b [a]*

Focusing on the upper atmosphere (0.05 mbar) of WASP-39b, the Titan-like and water-world-like haze cases again show a similar haze distribution, resembling the passive haze case. Figs. 12(iv and vii) show that haze is distributed from mid- to high-latitudes, concentrating around the edge of the superrotating jet, reaching maximum

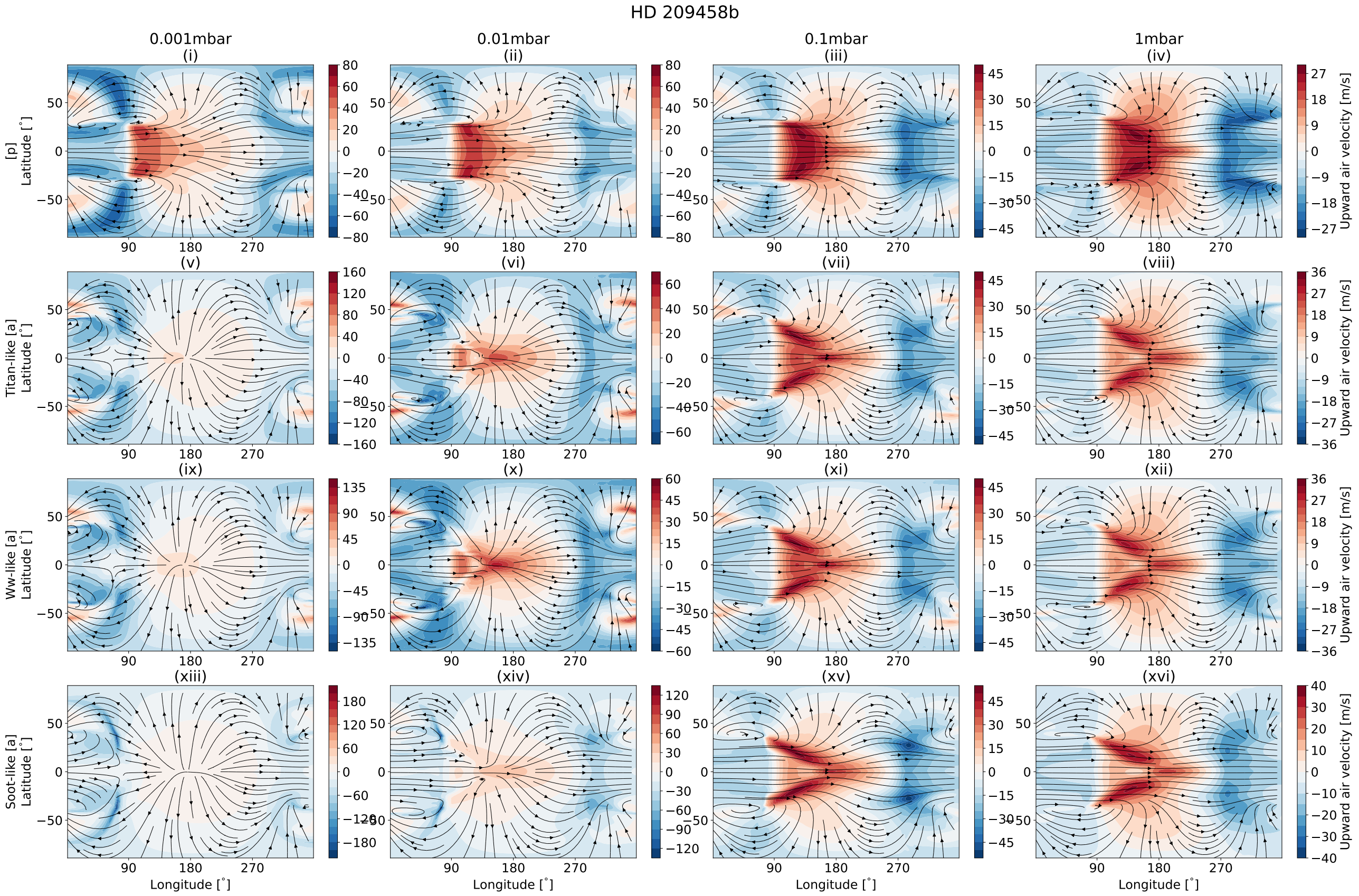

**Figure 10.** Vertical air velocity (contours) and with horizontal wind (arrows) from simulations of HD 209458b, following the same format of Fig. 9.

MMR within the nightside vortices. The haze distribution over the nightside vortices is less localised compared to the passive haze case. Instead of being tightly confined (see Fig. 12(i)), the haze is more broadly dispersed around the nightside vortices. It is also relatively depleted from the equatorial region in these two active haze cases, similar to the passive haze case. For the soot-like haze case, we note that the simulation again exhibits slight fluctuations in the haze distribution, demonstrated by the slight asymmetry present in Figs. 12(x–xii). Focusing on a pressure of 0.05 mbar, the haze distribution becomes nearly homogenised across longitudes and is not observed in all other cases and planets presented in this paper. There is slightly more haze around the edge of the jet and the nightside vortices. However the distribution over the night side vortices is narrowly confined over latitudes (see Fig. 12(x)).

As the pressure increases, hazes from the Titan-like and water-world-like simulations continue to peak over the nightside vortices, similar to the passive haze case, and the active haze cases from HD 189733b and HD 209458b. Part of the haze remains around the edge of the jet, extending to higher latitudes. For the soot-like haze case, the haze distribution between pressure 0.05–0.1 mbar remains similar. At a pressure of 1 mbar, the haze distribution again peaks over the nightside vortices, with the equatorial region depleted of haze. This is similar to other active haze cases and the passive haze case in WASP-39b.

Figs. 13(iv–vii) show that at a pressure of 0.05 mbar, the vertical velocity from our simulation with Titan-like and water-world-like haze is at maximum near the nightside vortices. The overall vertical wind distribution remains similar, with the equatorial region presenting a weak velocity compared to higher latitudes, on both day and night side. Similar to the passive haze case, this causes haze particles to sink down to the deeper atmosphere, reducing the MMR over the equator (see Figs. 12(iv and vii)). Figs. 5(vi and ix) also show a strong, positive, eddy mass flux at the edge of the jet at ±40° latitude, explaining the concentration of haze around the jet edge, similar to the passive haze case. After decomposing the wind and comparing the difference in wind strength with the passive haze case, Figs. 14(vi and ix) show a stronger eddies extending to high-latitudes over the nightside vortices, compared with weaker cyclonic eddies in the passive haze case. The strong eddies facilitate the accumulation of haze over a broader area surrounding the nightside vortices, resulting in the pattern observed in the simulation. For the soot-like haze case at a pressure of 0.05 mbar, Fig. 13(x) and 14(xii) do not show a clear structure of nightside vortices. Fig. 14(xii) also shows strong eddy component which facilitates the mixing between longitude, resulting in a nearly homogenised longitudinal distribution.

At higher pressures, the wind field of the Titan-like and water-world-like active haze cases remain similar. However, the eddy mass flux becomes weaker, resulting in the haze being trapped mostly around the nightside vortices. This resembles what is observed for HD 189733b in the deeper atmosphere where the eddy mass flux becomes weaker for the two active haze cases (see Figs. 6(vi and ix)). For the soot-like haze case, the wind field remains similar at a pressure of 0.1 mbar. However, at a pressure of 1 mbar, the wind structure resembles the other active haze cases, which results in a similar haze distribution.

In summary, in all our simulations with active haze, its radiative impact changes the thermal structure of the planetary atmosphere drastically, introducing isothermal regions and thermal inversions into the thermal structures. The haze MMR and its spatial distribution differ for each case due to the varying radiative forcing from haze altering the strength and the location of vertical and horizontal winds differently. Our results show that the stronger the absorption strength of the haze, the stronger the jet, and the smaller the difference in the day-to-night haze distribution. However, even under the influence of different radiative impacts, the haze distributions are still determined by the superrotating jet, the eddy mass flux and the decomposed wind in the same way as in the passive case, summarised in Sec. 3.1. More importantly, small-particle haze in most of the cases concentrates over the nightside vortices at pressure ≥0.1 mbar. This creates a stronger opacity source over the morning terminator, creating observational limb asymmetry in the transmission spectra (see discussion in Sec. 3.3.2).

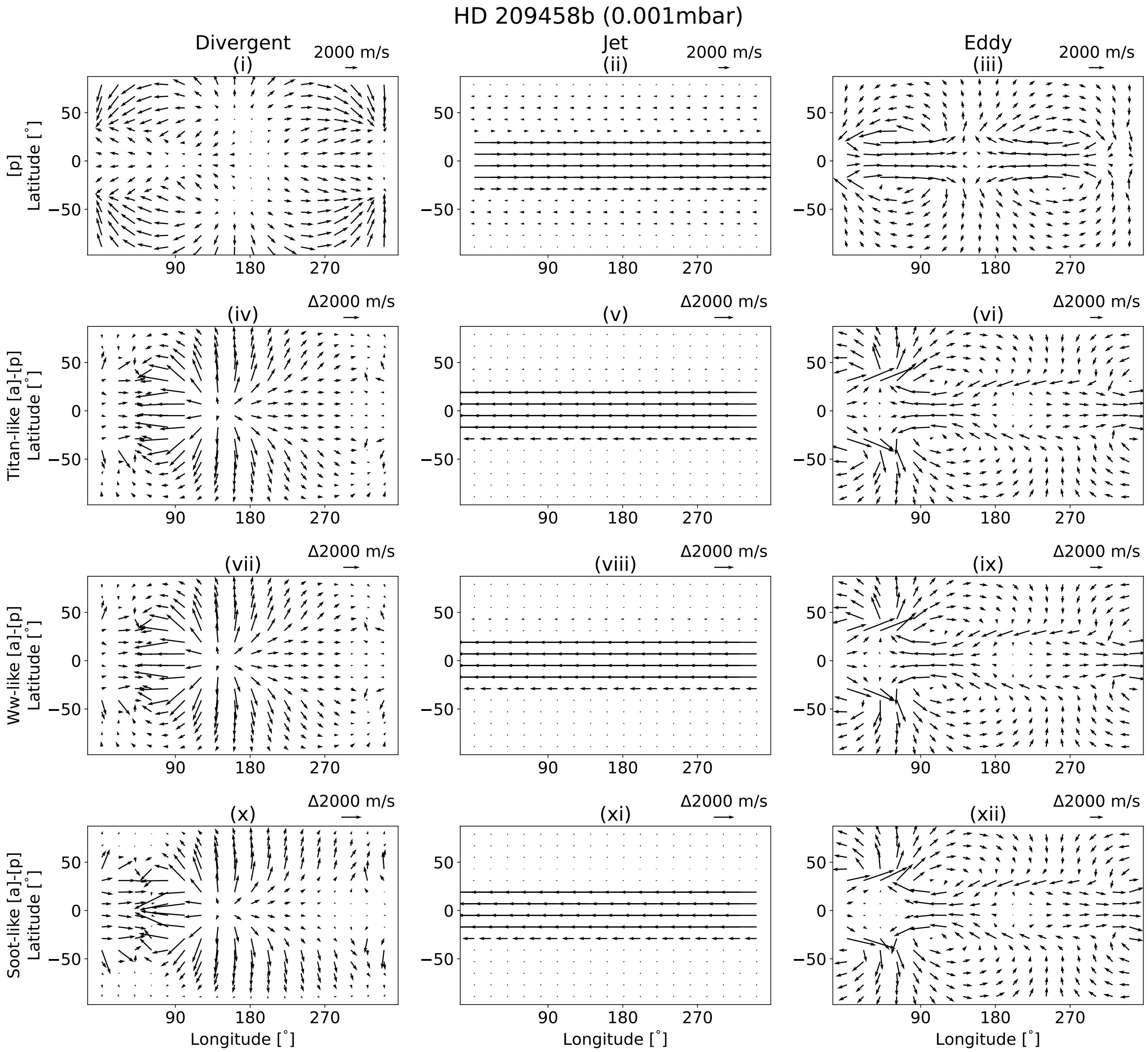

**Figure 11.** Components of the horizontal wind at a pressure of 0.001 mbar for all cases of HD 209458b, following the format of Fig. 8.

**Table 3.** The maximum zonal wind speed [m s$^{-1}$] of the superrotating jet. The pressure level [mbar] at which the peak velocity locates is included in bracket. "Ww" stands for water-world.

| | HD 189733 | HD 209458 | WASP-39 |
|---|---|---|---|
| Titan-like [a] | 7050.5 (0.90) | 5693.5 (0.16) | 4782.8 (0.20) |
| Ww-like [a] | 6846.9 (1.15) | 5671.8 (0.16) | 4698.5 (0.31) |
| Soot-like [a] | 7484.0 (10.17) | 6793.8 (1.41) | 6837.8 (0.33) |

### 3.3 Transmission Spectra

The implications of haze on observations can be explored through synthetic transmission spectra generated for our simulations using the UM (as first described in Lines et al. 2018). Examining the transmission spectra generated separately for the evening and morning terminators could potentially provide constraints on the atmospheric dynamics and relative timescales of mixing and chemical processes (Kempton et al. 2017). Previous work has already highlighted the potential for differences in atmospheric scale height (Fortney et al. 2010; Falco et al. 2024), clouds (Powell et al. 2018; Christie et al. 2021) and gas phase chemistry (Zamyatina et al. 2023, 2024) between the two limbs of hot-Jupiter atmospheres. Additionally, initial detections of such an asymmetry have been made with JWST for WASP-39b (Espinoza et al. 2024). In this section we calculate transmission spectra for our simulations using a high resolution 498-band spectral files, covering roughly the same wavelength range of 0.2–323$\mu$m (32 bands are used in the climate simulations–see Sec. 2.3). We show that soot-like haze produces the largest transit depth in the full synthetic transmission spectra, muting most spectral features, but create a flatter slope in the shortwave regime compared to the Titan-like and water-world-like haze. Then, we present synthetic transmission spectra over the two terminators, showing that the asymmetry in the limb depth is determined by the balance between the temperature and the haze opacity over the two limbs.

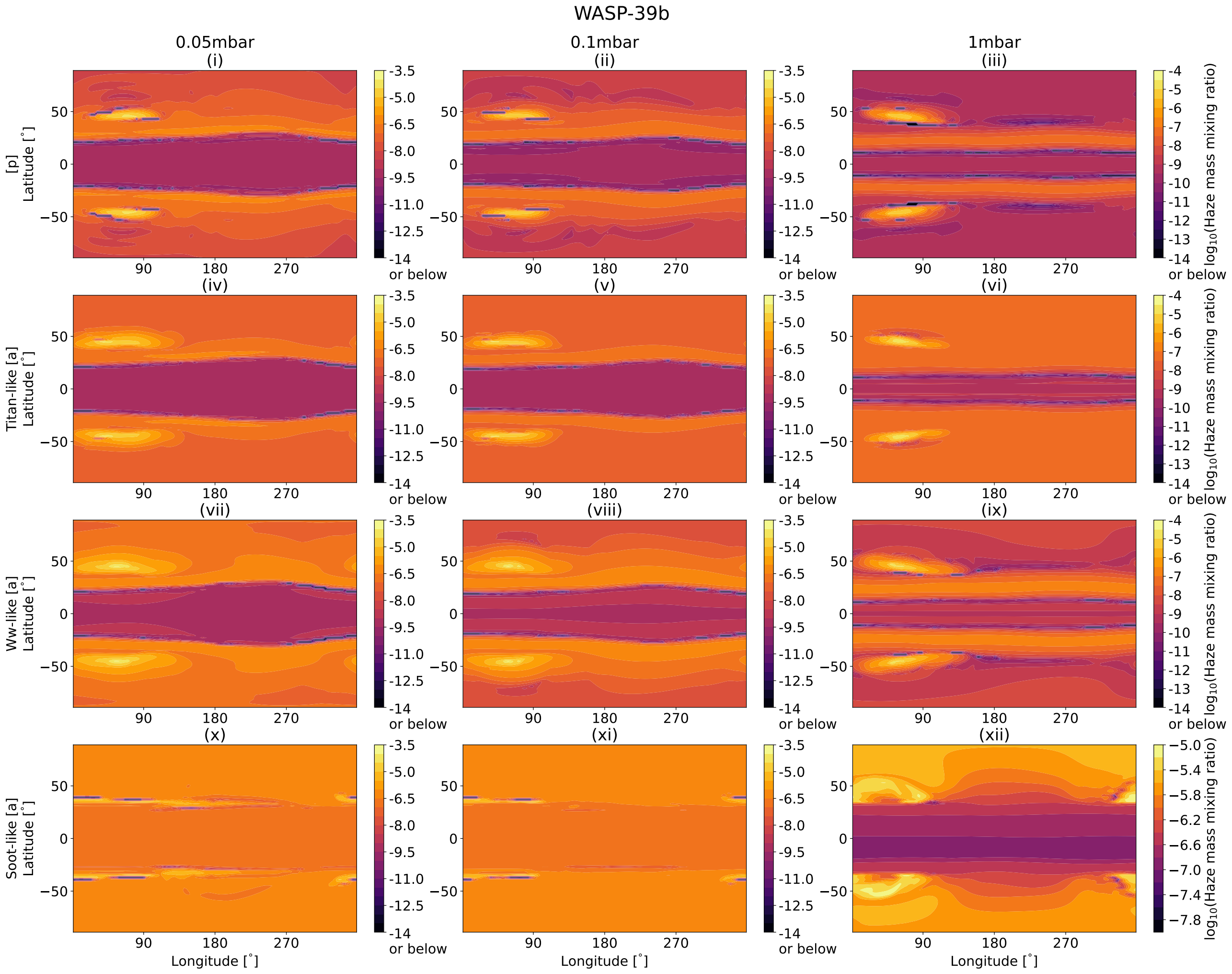

**Figure 12.** Mass mixing ratio of haze of WASP-39b for all cases at three isobaric surfaces, namely pressures of 0.02, 0.1 and 1 mbar. The haze mass mixing ratio is expressed in $\log_{10}$ scale. Note that the colour scales are different between different panels. "[p]" and "[a]" represents the passive and active haze case, respectively. "Ww" stands for water-world.

### 3.3.1 *Full Transmission Spectra*

Fig. 18 shows the full and individual limb transmission spectra for all of our simulations, including when the atmosphere is haze-free. List of gas species included in the simulations can be found in Sec. 2.3. The simulated spectra of HD 189733b are plotted alongside data from Swain et al. (2008),Pont et al. (2013),Sing et al. (2016) and Fu et al. (2024), HD 209458b alongside data from Evans et al. (2015) and Sing et al. (2016) and WASP-39b alongside data from Nikolov et al. (2016),Wakeford et al. (2018),Carter et al. (2024),Powell et al. (2024) and Espinoza et al. (2024). For each planet, the transmission spectra from the haze-free atmosphere are adjusted linearly (vertically with an additive offset) to match the water feature between 0.9–1.2 $\mu$m in the observational data. The active haze cases for the corresponding planets are then adjusted the same value as that of the haze-free case.

We note that previous work which examines the impact of haze at short wavelengths covers a wavelength range of up to 0.3$\mu$m (see Ohno & Kawashima 2020; Steinrueck et al. 2023, for example). However in this work, the Titan-like and water-world-like haze optical property profiles are a combination of two sets of data, joined at 0.4$\mu$m (see Sec. 2.2 for details). This results in a sharp transition in the optical properties at 0.4$\mu$m (see Figs. 1(i–ii)), which is also reflected as an abrupt change in the synthetic transmission spectra at the same wavelength. As we have used the optimum, available, input data for this work we simply note this feature and focus on the general trends of the spectra.

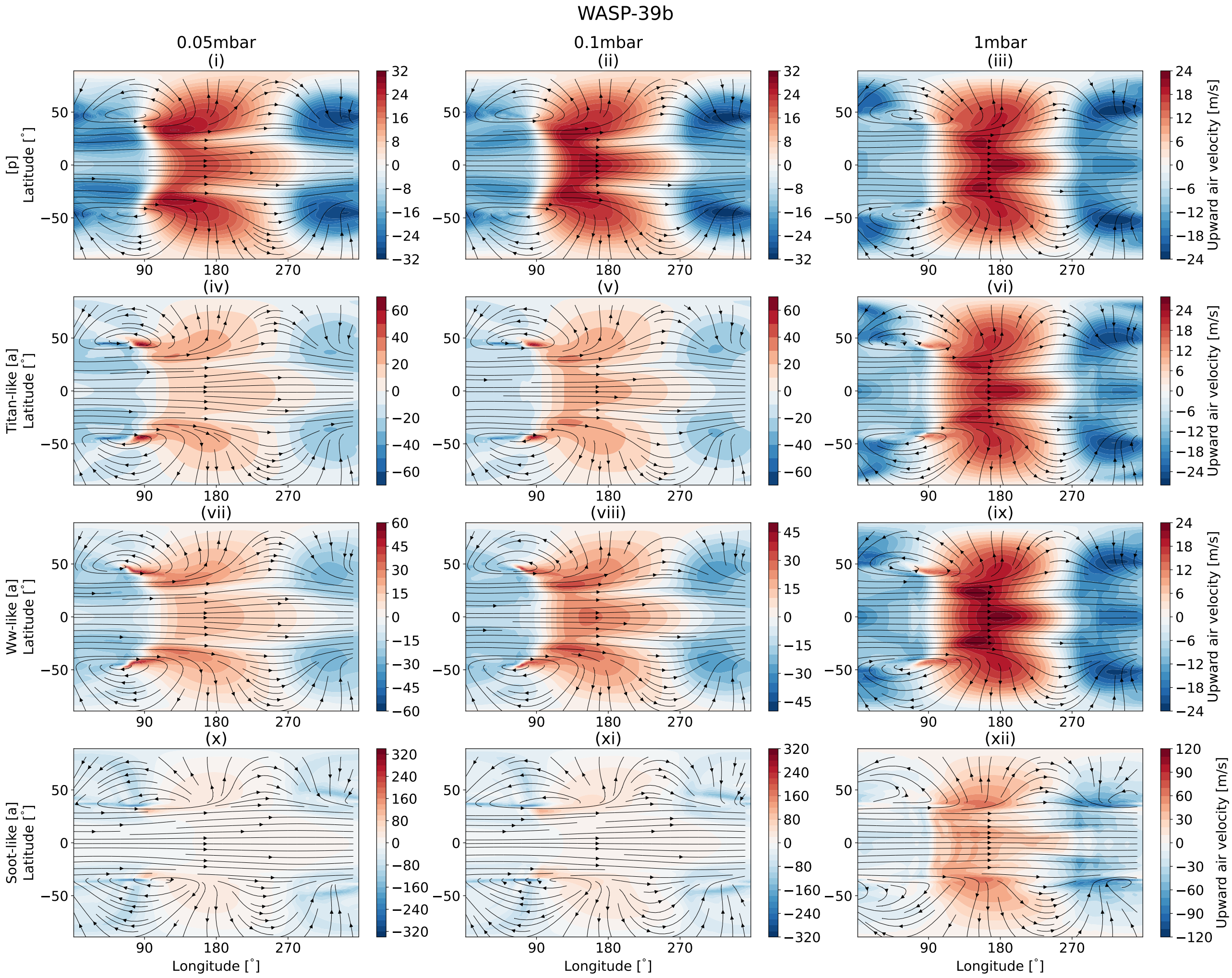

**Figure 13.** Vertical air velocity (contours) and with horizontal wind (arrows) from simulations of WASP-39b, following the format of Fig. 12.

Figs. 18(i, iii and v) show that for all planets, the synthetic spectra from the active haze cases are shifted upward, relative to the haze-free cases. The stronger the absorption strength of the haze, the larger the transit depth. This is due to the strong heating from the haze, producing a hotter thermal structure (see Figs. 17(i–iii)) which results in the atmosphere being "puffier" with a larger scale height when compared to the haze-free cases. The spectra produced under the presence of Titan-like and water-world-like haze resemble each other, due to the similar optical profiles creating a similar thermal structure (see Figs. 17(i–iii) and discussions from Sec. 3.2).

Figs. 18(i and v) show that for HD 189733b and WASP-39b, most of the spectral features are still preserved in the Titan-like and water-world-like haze cases compared to the haze-free cases. This is due to these two active hazes being optically thinner with a smaller extinction efficiency and absorbing lesser radiation (see Fig. 1), therefore retaining stronger spectral signatures. On the contrary, soot-like haze reduces the intensity of most of the spectral features at the shorter wavelengths. This is due to the optical properties of the soot-like haze exhibiting a grey structure (i.e., independent of the wavelength) with the strongest extinction efficiency, muting the spectral features across the shortwave regime. Towards longer wavelengths, the spectra begins to show the spectral features again because the extinction efficiency for all haze types are also relatively weak at longer wavelengths (see Fig. 1). The gas opacity at these wavelengths is of similar or greater strength than the haze opacity. For HD 209458b, the spectral features are still visible in all cases with little differences between the active and haze-free cases (see Fig. 18(iii)). This is unsurprising, as the active haze cases for this planet show a smaller haze MMR (see Fig. 3(ii)), in turn, leading to a lower haze opacity and the gas opacity dominating.

For all planets, Figs. 18(i, iii and v) also show that the slope of the spectra in the UV to optical regime is steeper than the Rayleigh slope under the presence of Titan-like and water-world-like haze. This is due to the extinction efficiency decreasing drastically from 0.3–1 $\mu$m (see Fig. 1), hence creating a much steeper slope as the wavelength transitions from UV to optical regime (Ohno & Kawashima 2020). The UV slope induced by the soot-like haze, on the other hand, is flatter. This is due to the grey optical profile which reduces the opacity gradient, resulting in a similar opacity at each pressure level.

Given our model parameters (i.e., particle radius of 1.5 nm and a haze production rate of $1{\times}10^{-12}$ kg m$^{-2}$ s$^{-1}$ – see Sec. 2.1) and comparing with the observational data, our results suggest that for HD 189733b, Titan-like or water-world-like haze fit the observational data better. To preserve the spectral features, the potential haze opacity source in the atmosphere of HD 189733b requires a weaker extinction efficiency than soot-like haze, and similar to Titan-like and water-world-like haze, or it might be located deeper in the atmosphere than our simulations setup. To create the steep UV slope as seen in the observational data, the optical profile of haze needs to show a rapid decrease in extinction coefficient between the UV and optical regime, similar to the profile of Titan-like and water-world-like haze (see Fig. 1). For HD 209458b, our results are inconclusive in determining whether it exhibits the presence of haze in its atmosphere as the spectra with and without haze are very similar. For WASP-39b, our haze-free case and the Titan-like and water-world-like haze fit the observations, therefore it is also inconclusive in determining whether it exhibits the presence of haze. However, if haze is assumed to be present in the atmosphere of HD 209458b and WASP-39b, the potential haze opacity source is likely to have a weaker extinction efficiency than soot-like haze, or is located in the deeper atmosphere, in order to preserve the spectral features. For WASP-39b, the potential haze would also require a grey optical profile to weaken the opacity gradient so as to create a flat UV slope. Again, we note that our discussions here are based on the model parameters adopted in this work. A lower haze production rate $F_0$, for example in the order of $10^{-15}$–$10^{-16}$ kg m$^{-2}$ s$^{-1}$ suggested by Arfaux & Lavvas (2022, 2023, 2024) in the study of WASP-39b, could also lead to a weaker haze opacity source and preservation of the spectral features.

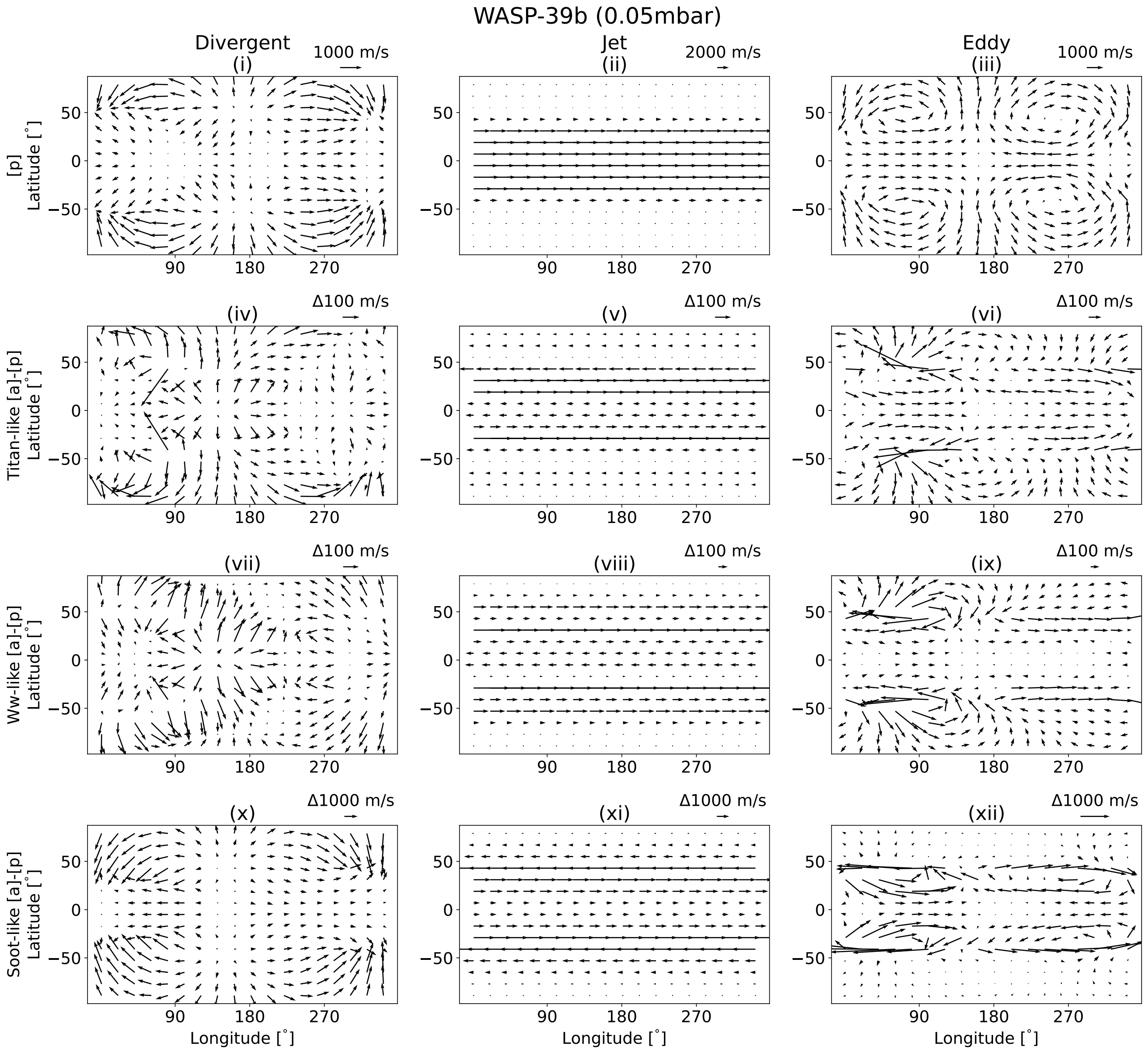

**Figure 14.** Components of the horizontal wind at a pressure of 0.001 mbar for all cases of WASP-39b, following the format of Fig. 11.

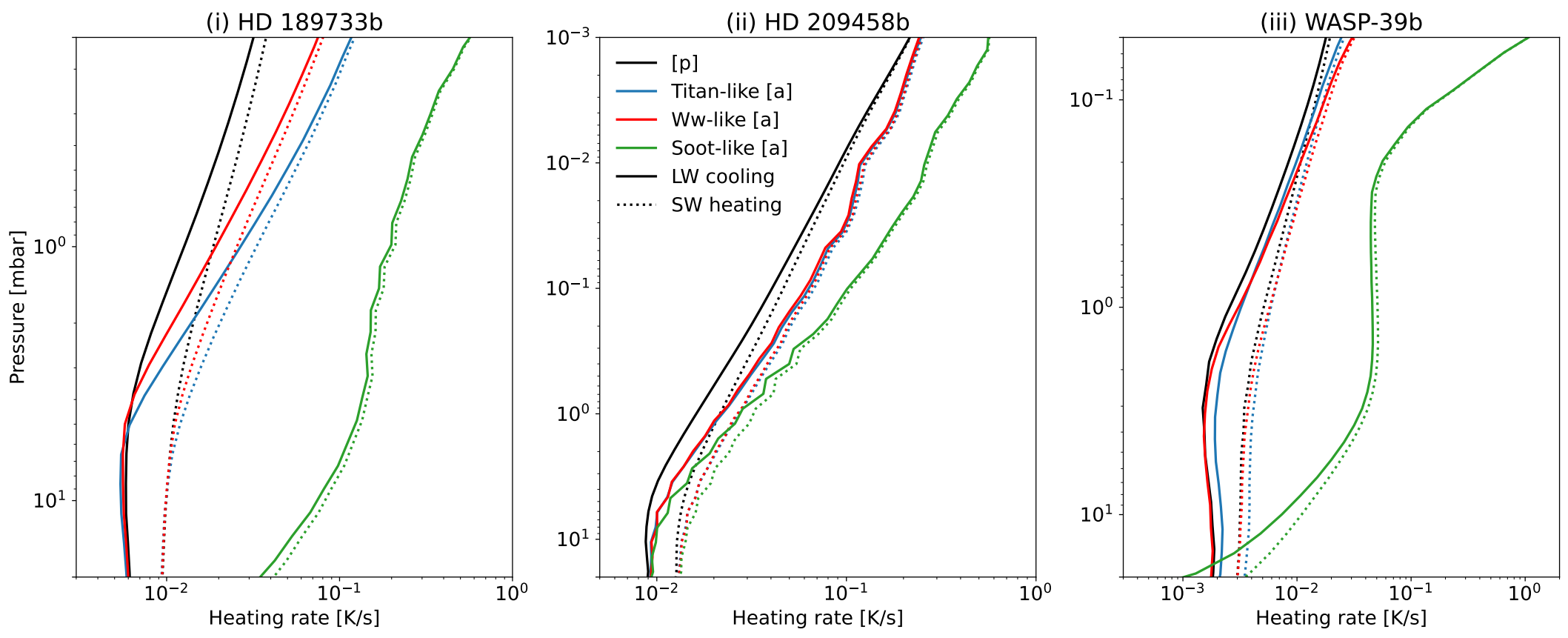

**Figure 15.** Dayside-mean shortwave (SW) heating rate (dotted) and longwave (LW) cooling rate (solid) for all cases of HD 189733b, HD 209458b and WASP-39b. "[p]" and "[a]" represents the passive and active haze case, respectively. "Ww" stands for water-world.

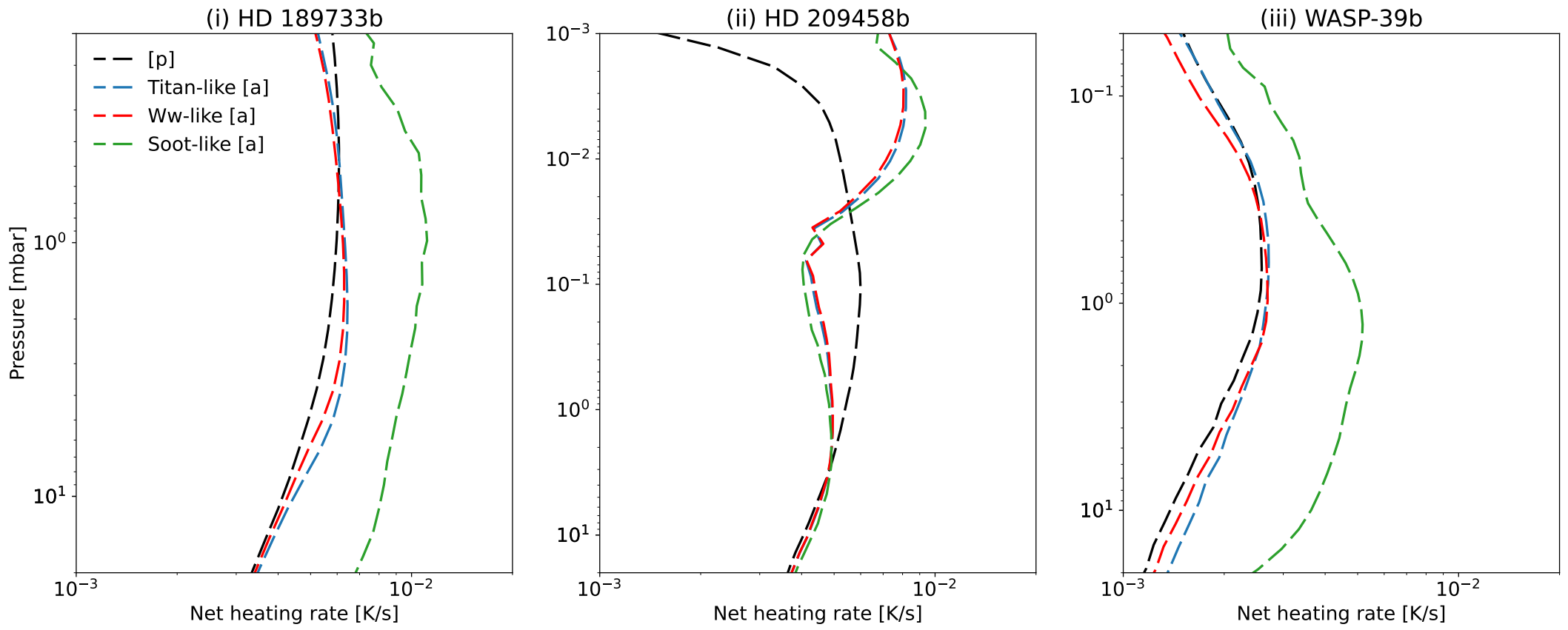

**Figure 16.** Dayside-mean net heating rate for all cases of HD 189733b, HD 209458b and WASP-39b. "[p]" and "[a]" represents the passive and active haze case, respectively. "Ww" stands for water-world.

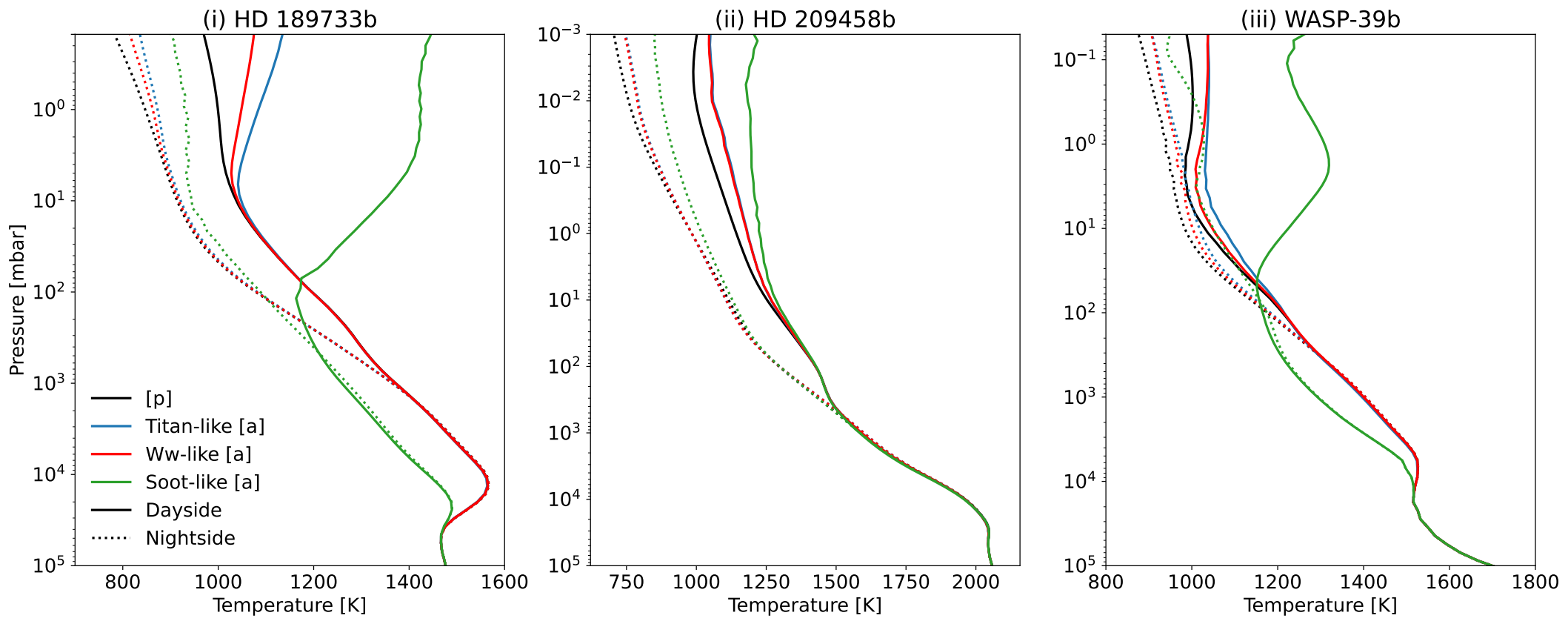

**Figure 17.** Dayside- (solid) and nightside-mean (dotted) thermal structures for all cases of HD 189733b, HD 209458b and WASP-39b. "[p]" and "[a]" represents the passive and active haze case, respectively. "Ww" stands for water-world.

### 3.3.2 *Limb Asymmetry*

The transmission spectra between the two terminators will differ as there are clear asymmetries in the structure of the dynamics and wind patterns, through the jet widths and eddies structures, and more significantly the positions of the nightside vortices (see e.g., Figs. 7, 10 and 13). These result in an asymmetrical distribution of temperature and haze MMR between the evening and morning terminators in all our simulations to varying degrees, therefore creating different temperature structures and haze opacities over the two limbs (see e.g., Figs. 6, 9 and 12). To understand the difference of transmission spectra between the two limbs, it is vital to understand how the thermal structure and haze MMR differ over the two terminators, as both these elements will have an impact on the measured radius and, thereby, directly impact the transmission spectrum.

Fig. 19 shows the morning (between longitude of 80° and 100°) and evening (between longitude of 260° and 280°) terminator-mean thermal structures. The thermal structure here in general resembles those shown in Figs. 17(i–iii). For all active haze cases, the evening terminator exhibits a hotter thermal structure than the morning ter-minator, due to the prograde superrotating jet transporting heat to the nightside through the evening terminator (see Sec. 3.2).

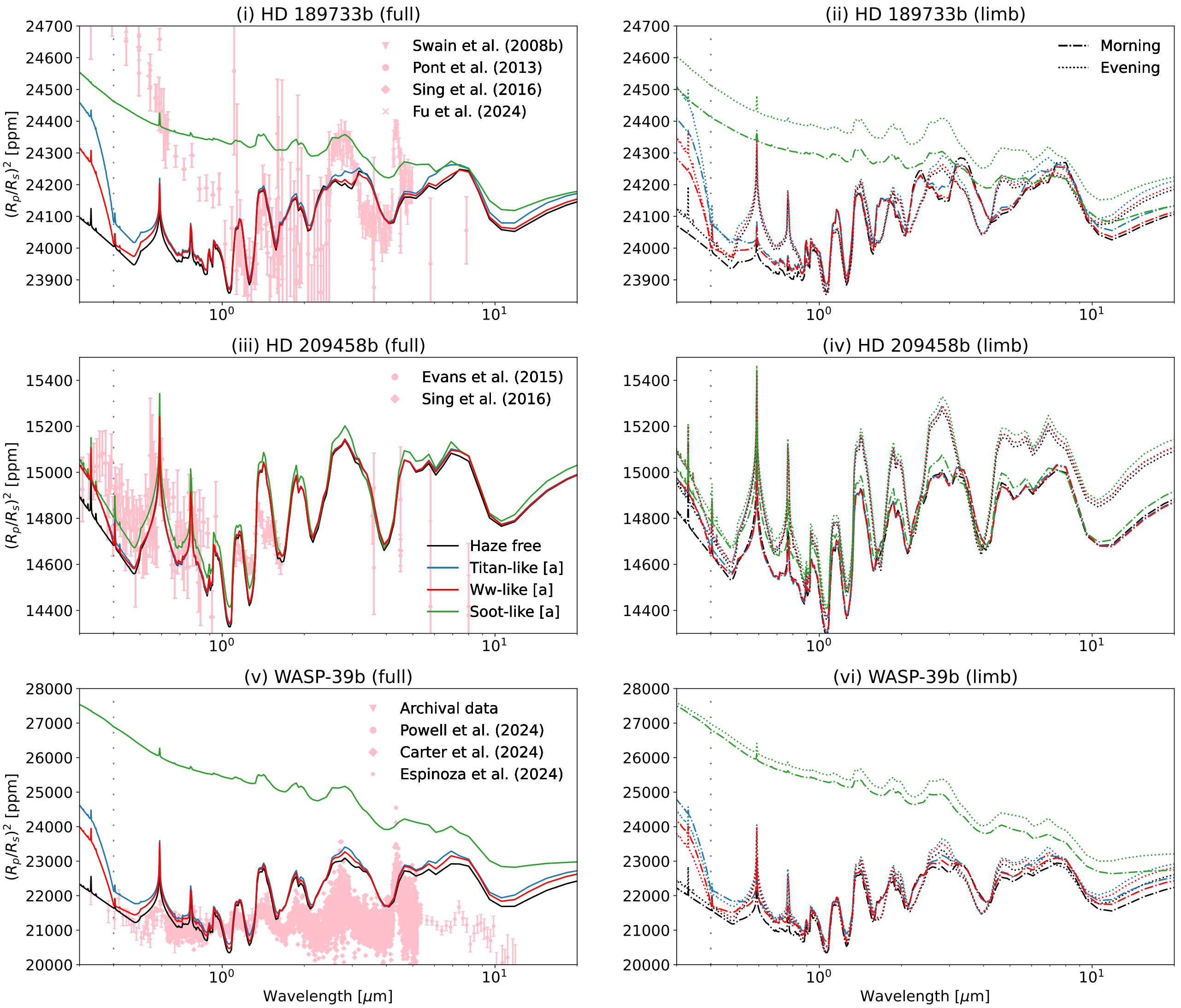

**Figure 18.** Transmission spectra of HD 189733b (i-ii), HD 209458b (iii–iv) and WASP-39b (v–vi) showing the full (1st column) and limb (2nd column) spectrum. "[a]" represents the active haze case. A vertical dotted line at 0.4 $\mu$m marks the wavelength where the discontinuity in the optical profile occurs in the Titan-like and water-world-like haze cases (see Sec. 3.3.1 for details). "Ww" stands for water-world.

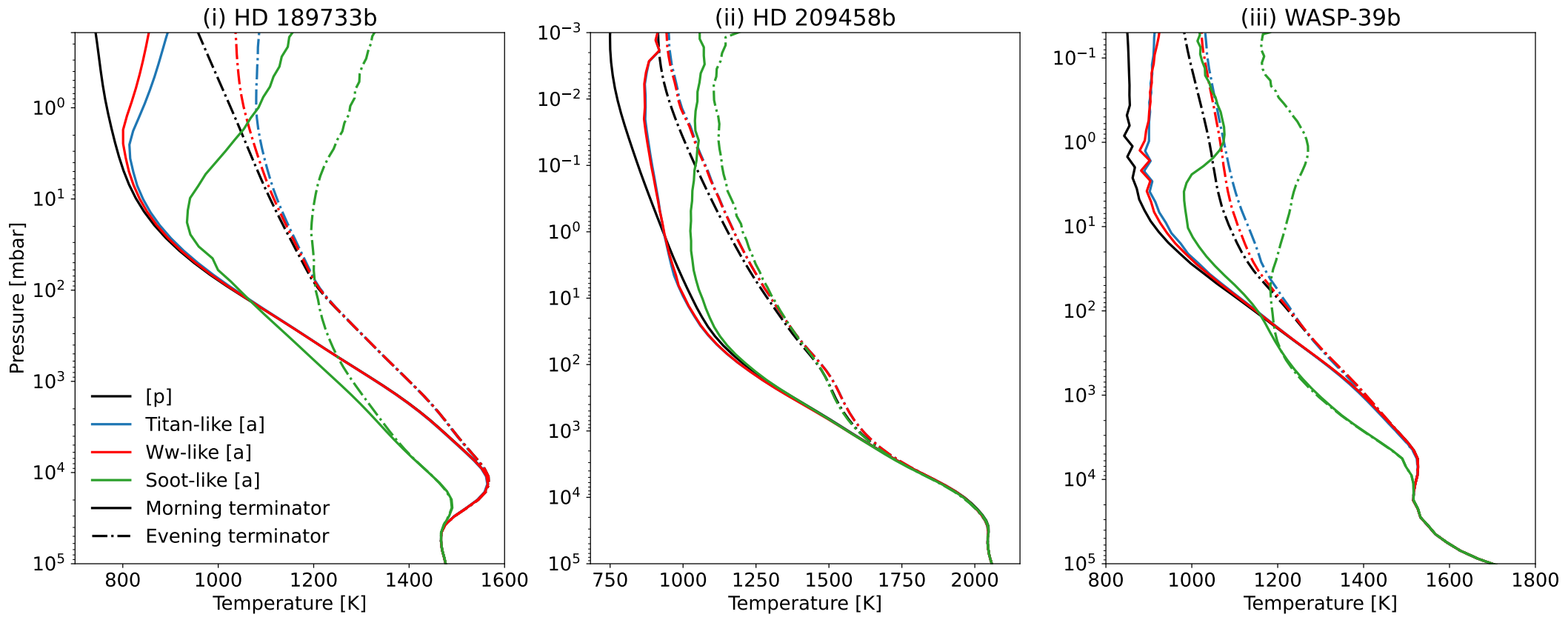

**Figure 19.** Morning (solid – between longitude of 80° and 100°) and evening (dash-dotted – between longitude of 260° and 280°) terminator-mean thermal structures for (i) HD 189733b, (ii) HD 209458b and (iii) WASP-39b. "[p]" and "[a]" represents the passive and active haze case, respectively. "Ww" stands for water-world.

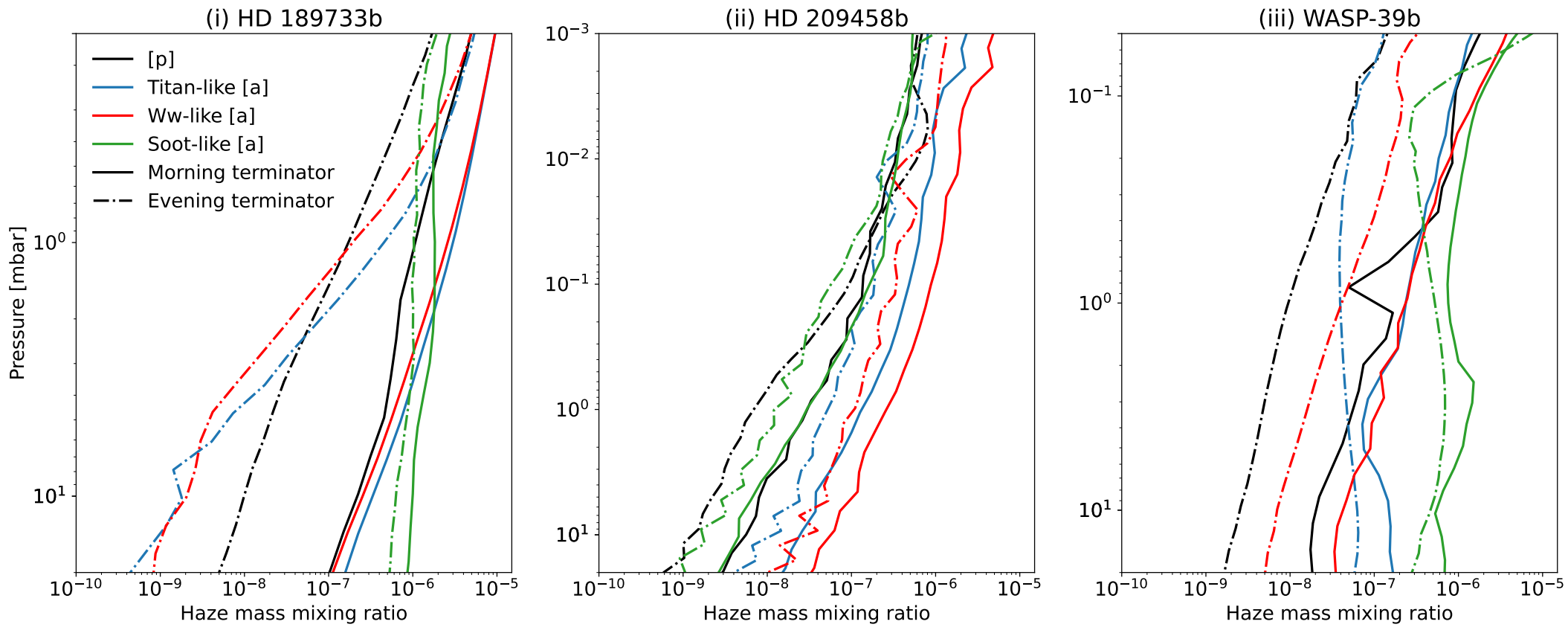

**Figure 20.** Morning and evening terminator-mean haze mass mixing ratio profiles, following the format of Fig. 19.

Fig. 20 shows the morning and evening terminator-mean of haze MMR profiles. For HD 189733b and HD 209458b, Figs. 20(i and ii) show that there is more haze MMR over the morning terminator, compared to the evening terminator in the upper atmosphere for all active haze cases. This is due to the fact that the nightside vortices in HD 189733b, which are located near the morning terminator, play a strong role in controlling the haze distribution. For HD 209458b, the haze distribution is mostly determined by the eddies as discussed in Sec. 3.2.2, which are focused over the morning terminator. The same result is seen in WASP-39b for all haze cases except for its soot-like haze case in which the evening terminator presents a slightly higher haze MMR at pressure of ∼0.05 mbar (see Fig. 12(x)) but the difference is almost negligible.

Figs. 18(ii, iv and vi) show the transmission spectra over the two limbs for all cases, including when the atmosphere is haze-free. The spectra are linearly adjusted the same way as for the full spectra (see Sec. 3.3.1). For both HD 189733b and HD 209458b in the Titan-like and water-world-like haze cases, Figs. 18(ii and iv) show that the evening terminator presents a larger transit depth than the morning terminator between the UV and optical wavelength regime. As discussed, Figs. 20(i and ii) show that there is more haze over the morning terminator which would introduce a stronger opacity, compared to the evening terminator in the upper atmosphere. However, a much hotter thermal structure over the evening terminator still results in a larger scale height (see Figs. 19(i and ii)). The increased opacity source over the morning terminator is not strong enough to raise the probing pressure level high enough and create a larger transit depth. In the long wavelength regime, the effect from haze is weaker. However, the temperature and pressure broadening of gas species are slightly different due to the different thermal and pressure structure, creating a larger transit depth over the morning terminator. For the soot-like haze case in both planets, the thermal structure over the evening terminator is so hot that it creates a larger transit depth across all wavelengths, regardless of the effect of haze and gas opacity.

For the WASP-39b Titan-like and water-world-like haze cases, Fig. 18(vi) shows that the morning limb also demonstrates a larger transit depth between the UV and optical wavelength range. This is opposite to the behaviour of the corresponding active haze cases for HD 189733b and HD 209458b. This feature is presented more clearly in Figs. 21(i, iii, v and vii), which show the limb transmission spectra of WASP-39b at different wavelengths, and plotted against JWST observational data from Espinoza et al. (2024). Each of our simulations here in Fig. 21 is linearly adjusted (vertically with an additive offset) to match the data from Espinoza et al. (2024). To investigate the balance between the temperature changes and haze opacity dependence on the limb transmission spectra for all active haze cases on WASP-39b, we calculate the limb transmission spectra with a clean-sky haze-free atmosphere. By excluding the optical effects of any haze, the resulting spectra is solely dependent on the underlying gas–phase structure and the resulting atmospheric thermal structure. To isolate the haze opacity dependence and obtain a “haze only” spectrum that shows the radiative effect of haze on the gas species, we calculate the difference in the spectra between an atmosphere with the inclusion of haze and a clean-sky atmosphere.

Fig. 22 shows the clean haze-free spectra (i, iv and vii), haze-only spectra (ii, v and viii) and the terminator difference in these two cases (iii, vi and ix). The terminator difference of the haze-free case is calculated as the spectra of the evening limb minus that of the morning limb (because the evening limb presents a larger transit depth). The terminator difference of the haze-only case is calculated as the spectra of the morning limb minus that of the evening limb (because the morning limb presents a larger transit depth). In other words, a larger terminator difference from the clean haze-free or the haze-only spectra would indicate the dominating contribution from temperature difference on the evening terminator, or haze opacity on the morning terminator accordingly from Figs. 22(iii, vi and ix). Figs. 22(i and iv) show that the difference over the two limbs due to the gas phase driven temperature differences is very little for the UV and optical wavelength regimes. However, Fig. 20(iii) shows that the morning terminator has a much higher haze MMR across all pressures. This results in a larger transit depth over the morning limb if we simply consider the impact from the strong haze opacity (see Figs. 22(ii and v)). The effect is particularly prominent in the shorter wavelength regime as this is where the haze layer has the highest opacity. At longer wavelength, the opacity of the haze layer weakens (see Sec. 3.3.1). Balancing the contribution from a hotter thermal structure over the evening limb and a stronger haze opacity over the morning limb, the latter becomes the dominating factor in creating a larger transit depth between the UV and optical wavelength. This can be seen from Figs. 22(iii and vi) in the UV–optical regime that wherever the haze opacity has a higher contribution over the gas-phase driven temperature differences, the morning terminator shows a larger transit depth (see Fig. 18(vi)). This is not seen in the Titan-like and water-world-like haze cases on HD 189733b and HD 209458b because the increase of haze MMR over the morning terminator compared with the evening terminator is much less at the pressure level probed by the transmission spectrum compared to the corresponding two haze cases in WASP-39b (see Fig. 20). Similarly for the soot-like haze case over WASP-39b, the haze distribution between the two terminators are similar (see Fig. 20(iii)). The additional haze opacity source over the morning limb is too little to raise a higher probing pressure level. In other words, in order to observe a larger transit depth between the UV and optical wavelength regime, the morning terminator would require a much higher haze concentration and a small temperature difference compared to the evening terminator. Although such feature is observed in the Titan-like and water-world-like haze case in WASP-39b, this might not be a typical feature for hot-Jupiters.

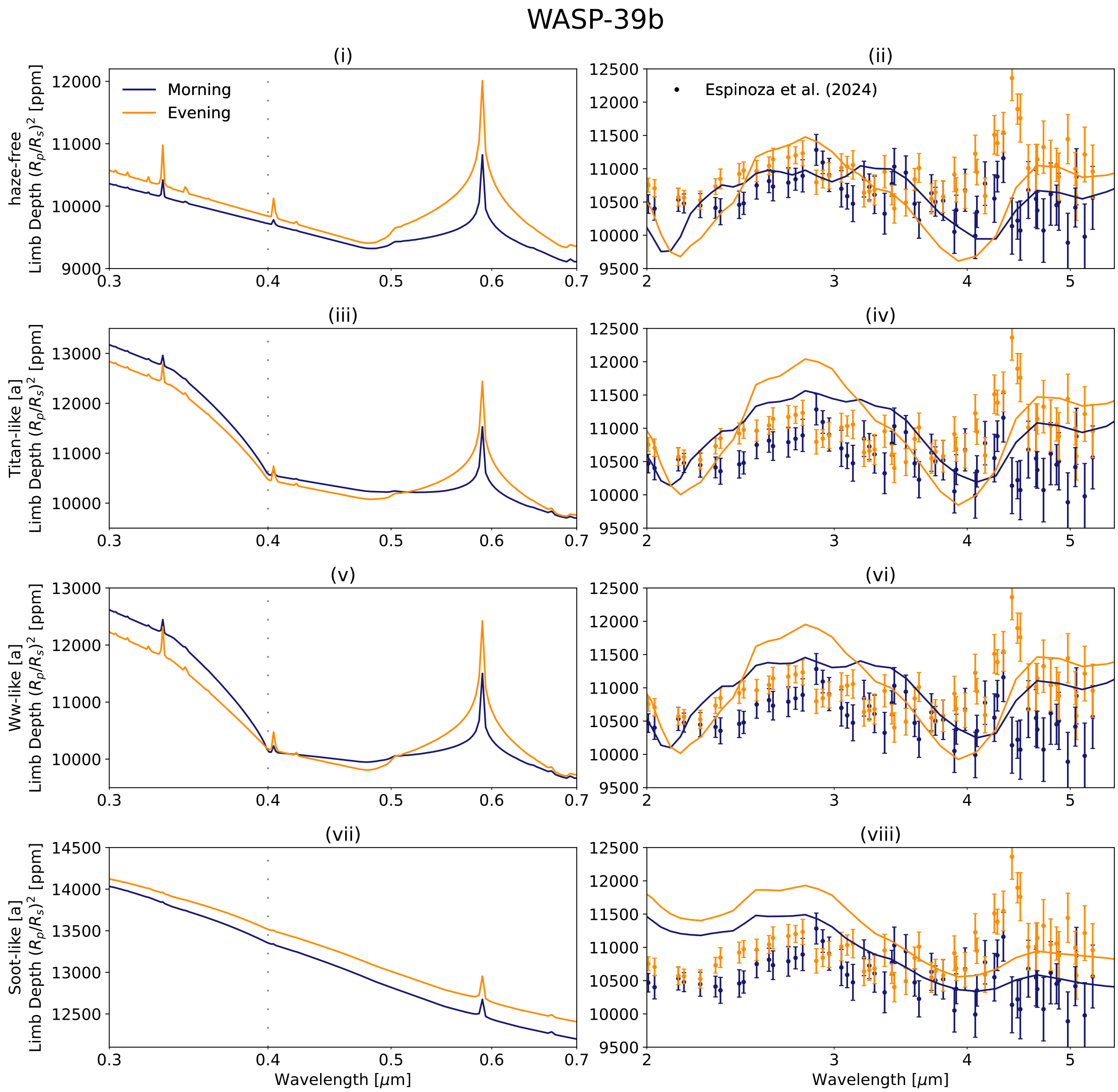

**Figure 21.** Limb transmission spectra for all cases of WASP-39b plotted against observations from Espinoza et al. (2024). Note that the scale is different in the 1st column. “[a]” represents the active haze case. A vertical dotted line at 0.4 $\mu$m marks the wavelength where the discontinuity in the optical profile occurs in the Titan-like and water-world-like haze cases (see Sec. 3.3.1 for details). “Ww” stands for water-world.

## 4 DISCUSSION

### 4.1 Haze Parameterisation

Following Steinrueck et al. (2021, 2023), our haze model described in Sec. 2.1 parameterizes haze production using a log-normal production rate against pressure, where here we have assumed a fixed mass haze production rate $F_0$. Ohno & Kawashima (2020) have found that a higher production rate would lead to an increased opacity in the atmosphere, resulting in a flatter slope in the UV–NIR regime in the transmission spectrum. In addition to that, Steinrueck et al. (2023) have demonstrated that more haze in the upper atmosphere would result in a stronger radiative heating and a faster wind, which would transport more haze particles away and reduce the concentration gradient in pressure. This is similar to our simulations where the water-world-like haze case shows a weaker jet strength, while the strongest absorber, soot-like haze, always presents the strongest jet.

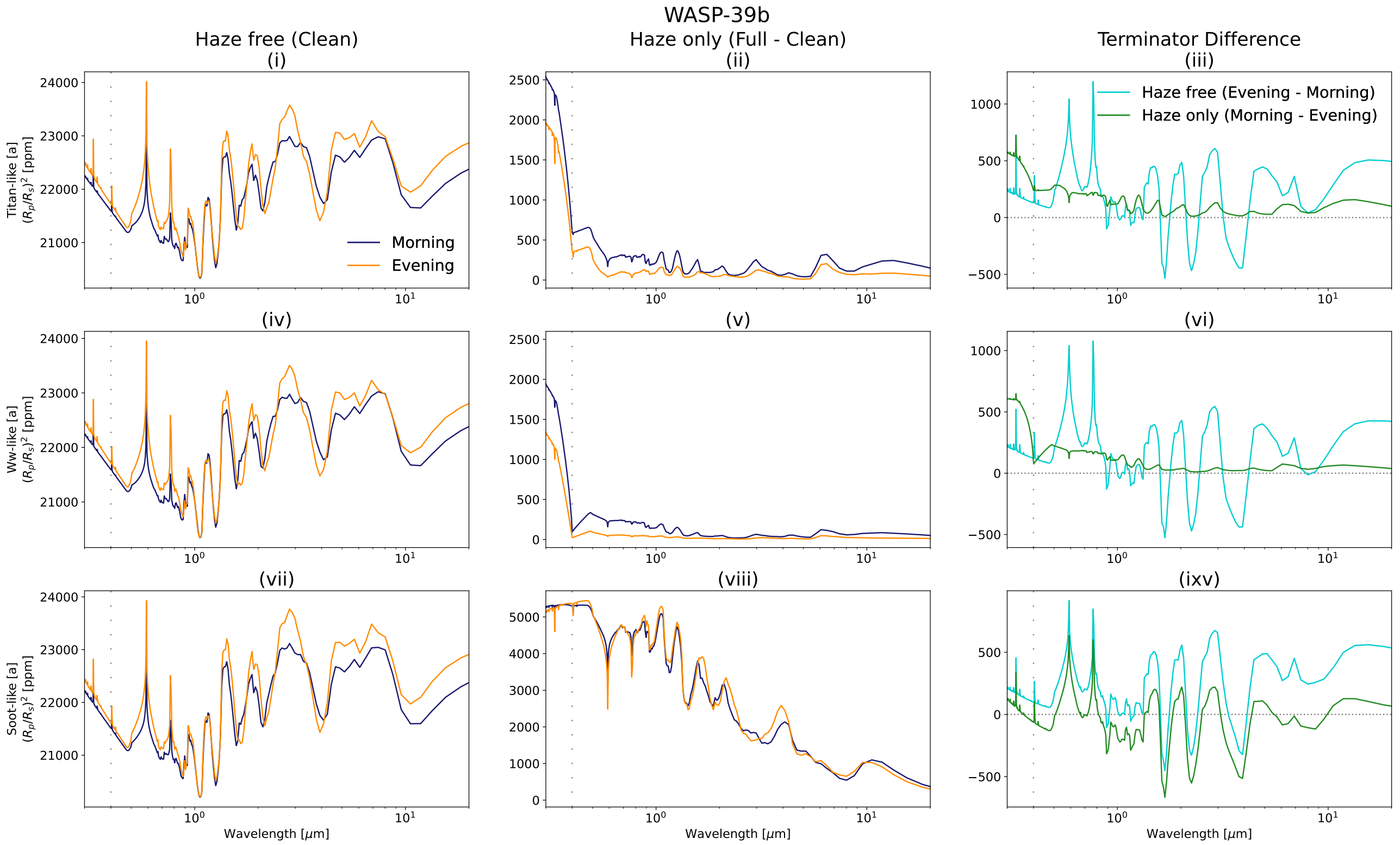

**Figure 22.** Limb transmission spectra of WASP-39b in all active haze cases. i, iv and vii: a haze-free atmosphere (clean spectrum); ii, v and viii: a hazy only atmosphere (full spectrum minus clean spectrum); iii, vi and ix: the terminator difference of the haze free case is calculated as the spectrum of the evening limb minus that of the morning limb. The haze-only case is calculated as the spectrum of the morning limb minus that of the evening limb. "[p]" and "[a]" represents the passive and active haze case, respectively. A vertical dotted line at 0.4 $\mu$m marks the wavelength where the discontinuity in the optical profile occurs in the Titan-like and water-world-like haze cases (see Sec. 3.3.1 for details). "Ww" stands for water-world.

However this work also shows that the wind field is changed due to the radiative forcing of the haze and we are unable to identify any trend in the haze concentration gradient in pressure.

Additionally, haze is produced through UV photolysis which depends on many factors, including temperature, stellar flux and the concentration of gas species. Haze production can be terminated if the haze layer is optically thick enough that it stops photons from entering the deeper atmosphere (Zerkle et al. 2012; Arney et al. 2016). A fixed production profile in this work might lead to overestimation of the haze MMR in the atmosphere. More importantly, if there are any changes to the chemical environment, such as departures from chemical equilibrium in the atmosphere at different rates between the morning and evening terminators due to quenching (Zamyatina et al. 2023), the haze production rate would change as the availability of its chemical ingredients would change. Zahnle et al. (2009) and Gao et al. (2020) have suggested that at equilibrium temperatures <1000 K, $CH_4$ will remain reactive but stable which favours the formation of haze. With a high temperature, $CH_4$ will be oxidized into CO and $CO_2$ by OH radicals, reducing the production of haze particles. However, Arfaux & Lavvas (2022) have also suggested that HCN can act as a main haze precursor up to 1300 K and CO for even hotter atmospheres (Hörst et al. 2018; He et al. 2019). For now, as previously noted, this work is intended as a first step to understand the radiative impact of haze on the atmospheric behaviours and observed spectra of different hot-Jupiters. Using a fixed production rate provides a simplified framework for isolating these effects from haze before introducing the full complexity of a chemically consistent network. However, future studies with a more complete 3D photochemistry module that incorporates the production of haze is needed to realistically and accurately capture the effect of haze in the atmosphere, alongside further laboratory experiments.

Regarding haze destruction, our model describes the combined effect of thermal destruction of haze and condensation of cloud species on the haze with a boundary condition only. Yet, in reality, the microphysical interaction between haze and cloud is complex. Haze acting as nucleation site facilitates cloud formation (Yu et al. 2021; Arfaux & Lavvas 2024). This mechanism can lower the concentration of haze particles and the condensing gas species in the atmosphere, reducing the intensity of the corresponding spectral features in the detected transmission spectrum. It can also increase cloud production with detectable features in observations, such as a flat spectrum (Arfaux & Lavvas 2024). The microphysical interactions between haze and cloud should be taken into account in a comprehensive 3D study of the impact of both haze and clouds in the planetary atmosphere.

## 4.2 Properties of Haze

We stress again that our here assumes haze particles to have a fixed radius of 1.5 nm, eliminating the possibility of coagulation and large-particles. This might lead to underestimation of the haze opacity, as the extinction strength increases as the radius of the particle increases. Steinrueck et al. (2021) have shown that small particles ≤30 nm can create a steep UV slope while increasing the particle size can flatten it. Adopting a radius of 1.5 nm in this work might be an acceptable assumption for the upper atmosphere where the particle size tends to be of the order of 1 nm (Lavvas & Arfaux 2021; Arfaux & Lavvas 2022) and the particle size has a weak effect on the atmospheric mixing (Steinrueck et al. 2021). However, for particle sizes ≥30 nm in the deeper atmosphere, the settling time might become an important factor dominating over the advection time, therefore changing the distribution pattern. For instance, Steinrueck et al. (2021) used the MITgcm with a grey radiative transfer treatment to simulate soot-like haze on HD 189733b and suggested that if haze is treated adopting a particle radius of ≥30 nm, there are in general higher MMR over the evening terminator than the morning terminator, which is the opposite to their previous results and our work here when haze is treated as smaller particles. Kempton et al. (2017) have also suggested that haze could settle out on the nightside and the evening terminator would therefore present a higher haze MMR. As a result, future work is needed to take into account coagulation and the modelling of particles of different sizes to accurately capture the settling and distribution of haze in the atmosphere.

We have also assumed haze particles to be spherical, rather than fractal agglomerates, and performed Mie calculation to obtain the optical properties of the haze. Yet, spherical haze and fractal agglomerates show a significant difference in their optical properties at different wavelengths. Wolf & Toon (2010) shows that haze particles, when treated as fractal agglomerates rather than spherical, have a higher absorbing strength in the shortwave regime, and a lower absorption strength in the longwave wavelength regime. Lodge et al. (2024) have showed that different calculation of haze optical properties, including Mie theory (Bohren & Huffman 2008, adopted in this work), discrete dipole approximation (DDA; Purcell & Pennypacker 1973; Draine & Flatau 1994) and modified mean-field theory (MMF; Berry & Percival 1986; Botet et al. 1997; Tazaki & Tanaka 2018), can yield different extinction strengths of the haze. For instance, soot-like haze, when treated as fractal agglomerates and their optical properties are calculated by DDA, will have a stronger absorption strength across most of the wavelengths, compared to when they are calculated using Mie theory or MMF, as opposed to what Wolf & Toon (2010) have suggested. These works show the sensitivity of the haze optical properties on their physical structure and calculation methods.

This study also explores the impact of Titan-like, water-world-like and soot-like haze in hot-Jupiters, showing that different haze optical profiles can affect the atmospheric circulation differently. More importantly, comparing to observational data and given our haze parametrization, Fig. 18 shows that potential hazes present in the atmosphere of HD 189733b and WASP-39b have optical profiles that lie between soot-like haze and Titan-like or water-world-like haze. With the equilibrium temperatures of hot-Jupiters ranging between 1000–2000 K, this could give rise to a haze with very different physical structures and optical properties. These potential hazes cannot be represented by the three haze types adopted in this work. With the lack of laboratory data for different haze types, this work highlights the urgency for measurements of optical properties of laboratory haze analogues to hot-Jupiters with different atmospheric composition, stellar types and haze formation environment.

### 4.3 Consequences of Model Choices

It is important to compare our simulations with those from other studies to ensure our predictions are robust. As discussed in Sec. 1, there is little 3D work which examines the impact of haze on hot-Jupiters. Part of our haze model is based on the setup of Steinrueck et al. (2023) which used the MITgcm to simulate Titan-like and soot-like hazes on HD 189733b but with a model top of <0.001 mbar while adopting different haze parameters with respect to our work (see Sec 2.1). Therefore, in this subsection we compare our results for HD 189733b to Steinrueck et al. (2023) and explore the sensitivity of our model choices on the resulting climate.

In Steinrueck et al. (2023), for their simulations with Titan-like haze, Steinrueck et al. (2023) adopted the optical properties of Lavvas et al. (2010) which based the real part ($n$) of the refractive indices on Khare et al. (1984) and retrieved the imaginary part (k) from the Descent Imager/Spectral Radiometer observations from the Huygens probe to Titan (Tomasko et al. 2008) (see their Fig. 1), which is slightly different from this work. Overall, the optical profiles between our work and Steinrueck et al. (2023) are similar, with the absorption strength of the Titan-like haze from Steinrueck et al. (2023) being much weaker between 0.7–3 $\mu$m, but slightly stronger between 0.4–0.7 and 3–30 $\mu$m than the optical profiles in this work.

Adopting an $F_0$ of 2.5×10$^{-11}$ kg m$^{-2}$ s$^{-1}$, Steinrueck et al. (2023) observed an increase of temperature of ∼280 K at ∼0.1 mbar. Compared to our simulation with $F_0$ of 1×10$^{-12}$ kg m$^{-2}$ s$^{-1}$ we see a smaller increase of ∼180 K on the day side. Both works show an almost isothermal temperature structure above ∼1 mbar. In terms of the haze distribution, both works also show that the Titan-like haze case has a higher haze mixing ratio than the soot-like haze. Steinrueck et al. (2023) show that at ∼0.1 mbar on the day side, the haze MMR reaches ∼1×10$^{-5}$, and a shortwave heating rate of almost ∼0.2 K s$^{-1}$. This is very similar to our results, where we see the haze MMR reaches ∼1.2×10$^{-5}$, and the shortwave heating rate almost ∼0.15 K s$^{-1}$. Such differences in haze MMR between the two studies could arise from different model choices that lead to variations in atmospheric circulation. Steinrueck et al. (2023) show that the haze particles are concentrated around the equatorial band between pressures of 0.1–1 mbar. Therefore, the haze will receive most of the stellar flux, exhibiting a strong shortwave heating. Whereas, in this work the equatorial region has a lower haze concentration compared to higher latitudes, and most haze is deposited around the nightside vortices (see Fig. 6(iv–vi)). The haze would receive a relatively lower amount of stellar flux, resulting in a slightly lower heating rate despite a higher haze MMR.

In Steinrueck et al. (2023), for their simulations with soot-like haze and a value of $F_0$ of 2.5×10$^{-11}$ kg m$^{-2}$ s$^{-1}$, there is an increase in temperature by ∼400 K on the day side at ∼0.1 mbar relative to their passive haze case. Compared to our simulation with $F_0$ of 1×10$^{-12}$ kg m$^{-2}$ s$^{-1}$, we see a larger increase of ∼500 K on the day side. Both works show that the day-to-night temperature difference with soot-like haze has increased significantly.

In terms of the haze distribution, at a pressure of ∼0.1 mbar, both Steinrueck et al. (2023) and our work here show that haze extends towards the equator and the morning terminator between the two hemispheres. The strength of the upwelling velocity also increases significantly in the soot-like haze case due to the thermal inversions in the atmosphere. Steinrueck et al. (2023) show an almost constant haze MMR profile on the day side at a pressure between 10–0.01 mbar, with an average value of ∼2×10$^{-6}$, reaching the shortwave heating rate of almost ∼0.3 K s$^{-1}$ at ∼0.1 mbar. From Fig. 3(i), we again show a higher haze MMR of ∼5×10$^{-6}$ at ∼0.1 mbar. As a result, despite both studies sharing a similar haze distribution, our work here shows a stronger dayside shortwave heating of ∼1.0 K s$^{-1}$ at ∼0.1 mbar due to a higher haze MMR (see Fig. 15(i)) compared to that of Steinrueck et al. (2023).

For the zonal jet structure, Steinrueck et al. (2023) show that with Titan-like haze, the superrotating jet is much stronger than when the atmosphere contains soot-like haze. This is because between pressures of 30–0.1 mbar, the atmosphere experiences a stronger radiative heating due to the optical properties of the Titan-like haze being wavelength dependent, allowing longer wavelength to penetrate through the atmosphere. This stronger heating in the mid atmosphere deposits additional energy to allow the formation of a stronger super-rotating jet (Koll & Komacek 2018, and see discussion in Sec. 3.2). In the soot-like haze case presented by Steinrueck et al. (2023), the net heating peaks at much lower pressures, rather than the mid atmosphere. In this pressure regime, the dynamical timescale is likely to be longer than the radiative timescale. As a result, the strong net heating in this low pressure regime does not effectively contribute to the formation of a stronger jet (Komacek & Showman 2016). Similar studies have also been conducted by Kataria et al. (2014) who varied the metallicity and the mean molecular weight in the atmosphere, and explored the formation of a superrotating jet at different pressure levels. Their work also shows that the formation of the jet depends on the energy budget and where the strongest heating rate is located in the atmosphere. Contrary to Steinrueck et al. (2023), our results show that a stronger jet is formed with soot-like haze in all planets (see Sec. 3.2, Fig. 4 and Tab. 3). This is due to soot-like haze presenting the strongest net heating in the deeper atmosphere (see Fig. 15 and Sec. 3.2). This drives the strongest jet, unlike what is observed in the case from Steinrueck et al. (2023). This could be due to the model dynamics which bring haze to different parts of the atmosphere, allowing the haze particles to receive different intensities of stellar flux. Steinrueck et al. (2023) show that their Titan-like haze particles are concentrated around the equatorial band on the day side at all pressure levels whereas their soot-like haze particles are relatively depleted on the day side between 0.01–0.1 mbar. Their Titan-like haze case also shows a higher haze MMR than that of the soot-like haze case. As a result, their Titan-like haze case exhibits stronger heating and drives a stronger jet. Yet, in this work the distribution patterns between our Titan-like and soot-like are similar and both cases receive a similar amount of stellar flux in the upper atmosphere, allowing the soot-like haze to present the strongest heating (see Fig. 6). This comparison highlights the sensitivity of haze transport due to model dynamics which would allow the haze particles to receive varying degrees of stellar flux, resulting in different net heating between the two models.

In summary, our work and that of Steinrueck et al. (2023) show differences in the atmospheric circulation on hot-Jupiters in the presence of haze. However, despite our work adopting a downward shift of median pressure, a lower value of $F_0$, and a lower outer atmospheric boundary below the peak haze production region at $\leq$0.1 mbar compared to Steinrueck et al. (2023), our results in general are in good agreement with theirs. Both works show similar haze concentrations and distributions, and that haze significantly heats up the atmosphere, changing the thermal structure drastically which in turn alters the haze distribution.

### 4.4 Implications for Observations

Our work agrees well with previous results which show that haze mutes spectral features (Morley et al. 2015; Ohno & Kawashima 2020; He et al. 2024), increases atmospheric scale height due to the radiative heating in the upper atmosphere (Lavvas & Arfaux 2021), as well as leads to a flatter UV slope for soot-like haze and a steeper slope for the Titan-like and water-world-like haze (Morley et al. 2015; Ohno & Kawashima 2020; Lavvas & Arfaux 2021; Steinrueck et al. 2021, 2023). However, our discussion here is based on the assumption that there is haze present in the atmosphere of these planets, with production parameters set by our haze model. We further assume that haze, along side gas species, are the sole factor impacting the transmission spectrum. In reality, clouds could also be present and would likely create a flat spectrum and mute the spectral signatures (Morley et al. 2015; Sing et al. 2016; Christie et al. 2021, 2022; Arfaux & Lavvas 2024; Espinoza et al. 2024). Powell et al. (2018) simulated a range of hot-Jupiters' atmospheres and concluded that those with equilibrium temperature $\leq$1700 K favour the formation of $TiO_2$ and $MgSiO_3$ clouds. Powell et al. (2018) further showed that cloud opacity can mute spectral features. Since the two terminators may exhibit different temperatures, each favouring cloud formation to different extent, observational limb asymmetry could also arise due to variation in cloud opacity between the two terminators. Even though Parmentier et al. (2016) have suggested that MnS clouds, rather than $MgSiO_3$ clouds, are more likely to form in planets with equilibrium temperature $\leq$1600 K, and that cloud composition could vary with different temperature, all of our target planets fall below this threshold, we therefore emphasise that clouds are likely to form in these planets. Moreover, stellar activity would impact the features observed in these spectra. For instance, McCullough et al. (2014) suggested un-occulted star spots could result in a spectral slope steeper than the Rayleigh scattering slope in the transmission spectrum of HD 189733b. Espinoza et al. (2019) have also suggested the effect of star spots to explain the steeper spectral slope from observations of WASP-19b. Although the impact of clouds and stellar activity lie outside the scope of this work, these factors could be spectrally degenerate with the effect of haze, producing indistinguishable features on the transmission spectra. Therefore, other than performing a photochemistry analysis to examine the likelihood of an atmosphere hosting haze, we recommend performing analysis on limb asymmetry between the UV and optical wavelength regime, using 3D GCM to capture the atmospheric dynamics accurately, which might help to better constraint the presence of haze.

As discussed in Sec. 3.3, asymmetry in transmission spectra over two limbs may enable the inference of the underlying atmospheric dynamics of the planet and be potentially detectable by JWST (Rustamkulov et al. 2022). From previous studies, neither can cloud opacity (Powell et al. 2018; Christie et al. 2021; Espinoza et al. 2024), disequilibrium thermal chemistry (Zamyatina et al. 2024) or a haze-free atmosphere (passive haze cases in this work–see Figs 22(i, iv and vii)) result in the morning terminator presenting a stronger signal than the evening terminator in the UV–optical regime. To create such an effect, the atmosphere would either need a retrograde jet which results in a hotter morning terminator (this can be caused by strong magnetic fields, Hindle et al. 2019), or a much stronger opacity source in the upper atmosphere of the morning limb which would raise the photosphere. This work satisfies the latter, showing that the asymmetry between the two limbs is balanced between the gas-phase driven temperature difference and the haze opacity (see Fig. 22). This is also in agreement with Steinrueck et al. (2021) who have demonstrated limb asymmetry and a larger transit depth over the morning limb in the synthetic transmission spectrum of HD 189733b by assuming haze particles to be passive, and with a particle size of 3 nm, using the MITgcm (note that their results did not include how haze would heat up the two terminators, which would increase the scale height over both limbs).

In summary, we suggest that a detection of a larger transit depth over the morning limb in the UV–optical regime might act as a strong indicator for the presence of haze, at least for small-particle sizes, for hot-Jupiters, as demonstrated by our simulations of WASP-39b (see Figs. 21 and 22). Focusing on such asymmetry in the shortwave regime also allow us to isolate the effect of haze, distinguishing it from that of thermal chemistry and gas-phase driven temperature difference, both of which have a strong effect at longwave regime. Here we stress again that the advection and settling time, hence haze distribution, might change when larger particles are considered (see further discussions in Sec. 4.2). A full haze study that examines haze production, physical and optical parameters, as well as various model choices is needed to explore their impact on limb asymmetry. Furthermore, although the atmospheric circulation of these tidally-synchronised hot-Jupiters tends to trap haze around the nightside vortices (see Sec. 3.2), resulting in a larger haze opacity over the morning limb, the evening limb could still present a stronger transit depth due to its high temperature, as seen in HD 189733b and HD 209458b. This asymmetry in morning limb dominating over the evening limb might not be a typical case among hot-Jupiters. Future studies are needed to examine the balance between haze opacity and the temperature difference between the two terminators across a wide population of hot-Jupiters. Understanding how much haze opacity is required over the morning limb to overcome the temperature difference will be essential for informing target selection, particularly for those aiming to use this method to test for the presence of haze.

#### *4.4.1 Observational Limb Asymmetry on WASP-39b*

As detections of limb asymmetry for WASP-39b have been made using JWST observations (Espinoza et al. 2024), Figs. 21(ii, iv, vi and viii) shows that from observational data, the morning terminator occasionally shows a stronger signal only between 3–4 $\mu$m and the fitted spectra from Espinoza et al. (2024) suggested a larger evening limb between 2~5 $\mu$m. Espinoza et al. (2024) simulated the impact of passive soot-like haze using the GCM SPARC/MITgcm (without considering the radiative feedback of haze on the thermal structure, but only the impact of haze opacity on the synthetic spectra). They assumed a haze particle radius of 30 nm and a production rate of $2.5\times10^{-12}$ kg m$^{-2}$ s$^{-1}$ (see Fig. 3c from Espinoza et al. 2024). Their results showed that haze has little effect on limb differences and the morning limb still produces a larger absorption in the $CH_4$ bands between 2.3–3.3 $\mu$m than the evening terminator due to the colder thermal structure over the morning limb, opposite to their fitted spectra. Espinoza et al. (2024) therefore suggested a relatively cloud-free evening terminator and a cloudy morning terminator, without the inclusion of haze, to match the observational data. However, as discussed in Secs. 3.2 and 3.3.2, haze can have a strong effect at altering the difference between the two terminators due to haze opacity raising the photosphere and the radiative heating from haze which increases the temperature in the upper atmosphere. Our soot-like haze case is in agreement with Espinoza et al. (2024)'s data, showing a hotter evening limb and a larger transit depth across all wavelengths. The difference between our simulations and Espinoza et al. (2024)'s could arise from the inclusion of radiative feedback on the thermal structure of the planetary atmosphere. Our work here shows that the soot-like haze can heat up the atmosphere drastically, creating a much hotter evening and morning terminators (see Fig. 19(iii)). This weakens the $CH_4$ features over both terminators due to a hotter thermal structure. In addition to a similar haze distribution on both terminators (see Fig. 20(iii), also see Sec. 3.3.2 for details), this results in the evening terminator still presenting a larger transit depth across all wavelengths. Our haze-free, Titan-like and water-world-like haze cases, on the other hand, align with the hazy simulations of Espinoza et al. (2024), demonstrating a weak effect of haze opacity at longer wavelengths and that they do not fit the observational data. Considering our simulation results with the observational data and the hazy simulations from Espinoza et al. (2024), we are unable to determine the haziness of WASP-39b. However, our work here highlights the importance of considering the radiative feedback of haze on the atmosphere when predicting observational limb asymmetry.

In the discussion of the haziness of WASP-39b, some observational data (Nikolov et al. 2016; Pinhas et al. 2019) and previous modelling work (Arfaux & Lavvas 2022, 2024) have suggested the potential inclusion of haze when fitting the simulation results with observational data. In Figs. 21(i, iii, v and vii), we explore the detectability of limb asymmetry in the UV–optical regime. As discussed in Sec. 3.3.2, other than the haze-free and soot-like haze case, the Titan-like and water-world-like haze cases are able to produce a larger transit depth over the morning terminator between 0.3–0.5 $\mu$m. In the soot-like haze case, although the transit depth of evening terminator is slightly larger, the muted spectral features at such short wavelength can still be a strong indicator of the potential presence of haze. We therefore suggest the analysis of observational limb asymmetry in the UV–optical regime to better constrain the presence of haze in WASP-39b. We also underscore the importance of expanding the library of haze analogues from laboratory experiment to better capture the effect of potential haze present in the atmosphere of WASP-39b.

## 5 CONCLUSIONS

In this work, we used a 3D GCM, the UM, to simulate the advection, settling and radiative impact of different haze types, with a particle radius of 1.5 nm, in the atmospheres of three hot-Jupiters, namely HD 189733b, HD 209458b and WASP-39b. We assume that the haze production follows a log-normal distribution profile and the haze destruction is determined via a boundary condition. We first assess the results when the haze is treated as a radiatively passive tracer. Then, we evaluate the simulations where the haze is considered radiatively active, following the optical properties of Titan-like, water-world-like, and soot-like haze. We find that the radiative forcing imposed by the haze drastically changes the thermal structure, thereby altering the atmospheric circulation. In particular, the radiative effect of the soot-like haze is the strongest, introducing thermal inversions and isothermal structures in the atmosphere. Whereas the radiative effect of the Titan-like and water-world-like haze is very similar due to the similarities of their optical profiles. We show that the stronger the absorption strength of the haze, the stronger the superrotating jet, lesser the difference of the day-to-night haze distribution, and larger the transit depth in the synthetic transmission spectrum. Haze with higher extinction efficiency also leads to muted spectral features. Yet, we note that similar effects could arise from the presence of clouds, which are not included in the current simulations. In all circumstances regardless of the specific radiative forcing imposed by the haze or the varying atmospheric circulation responses among different planets, the haze distribution is controlled by the same three mechanisms, 1) the superrotating jet largely determines the day-to-night haze distribution, 2) eddies drive the latitudinal haze distribution, and 3) the divergent and eddy component of the wind control the finer structure of the haze distribution.

This work also shows that due to the tidally-synchronised nature of hot-Jupiters, small-particle haze tends to be trapped over the nightside vortices. As a result, there is a strong haze opacity source over the morning terminator, altering the transit depth between the two terminators. For HD 189733b and HD 209458b, the evening terminator still produces a larger transit depth due to its high temperature. For WASP-39b, the morning terminator produces a larger transit depth than that of the evening terminator in the shortwave regime (see Fig. 21) because the haze opacity over the morning limb dominates over the gas-phase driven temperature difference. This feature is not seen under the impact of cloud opacity and wind-driven disequilibrium chemistry, suggesting that such detection might not be spectrally degenerate. Although this might not be a typical detection feature in hot-Jupiters, our finding suggests that if such limb asymmetry is observed in the UV–optical wavelength regime, it might act as a strong indicator for the presence of haze, at least for small-particle sizes, in the atmosphere. We emphasize again that our results might depend on our model setup, such as the choice of particle size, peak haze production pressure, and haze mass production rate (see Secs. 2.1, 4.1 and 4.2 for further details). Future study is needed to assess the sensitivity of results to these parameters. However, this study provides a general framework for understanding how haze impacts the atmospheric dynamics and observed spectral features of various hot-Jupiters, which may exhibit inherently different circulation patterns. This work offers a foundation for future investigations into the effect of haze on other such planets, and provides insights for future observing missions aimed at understanding the atmospheric compositions of gas giant exoplanets, opening up possibilities in other means of haze detection.

## 6 ACKNOWLEDGMENTS

We acknowledge funding from the Bell Burnell Graduate Scholarship Fund (grant number BB005), administered and managed by the Institute of Physics, which made this work possible. This work was supported by a UKRI Future Leaders Fellowship [grant number MR/T040866/1], a Science and Technology Facilities Council Consolidated Grant [ST/R000395/1], a Science and Technology Facilities Council astronomy observation and theory small award [ST/Y00261X/1], and the Leverhulme Trust through a research project grant [RPG-2020-82]. Material produced using Met Office Software. We acknowledge use of the Monsoon2 system, a collaborative facility supplied under the Joint Weather and Climate Research Programme, a strategic partnership between the Met Office and the Natural Environment Research Council. This work used the DiRAC Complexity system, operated by the University of Leicester IT Services, which forms part of the STFC DiRAC HPC Facility (www.dirac.ac.uk). This equipment is funded by BIS National E-Infrastructure capital grant ST/K000373/1 and STFC DiRAC Operations grant ST/K0003259/1. DiRAC is part of the National e-Infrastructure. M.S. acknowledges support from the 51 Pegasi b Fellowship, funded by the Heising-Simons Foundation.

## DATA AVAILABILITY

The research data supporting this publication are openly available with CC BY Mak (2025).